\documentclass[12pt,a4paper]{article}

\usepackage{amsmath,amssymb}
\usepackage[nosort]{cite}
\usepackage[hyperref,bulletsep]{collect}

\makeatletter
\let\old@startsection=\@startsection
\renewcommand{\@startsection}[6]{\old@startsection{#1}{#2}{#3}{#4}{#5}{#6\mathversion{bold}}}
\makeatother

\makeatletter \@addtoreset{equation}{section} \makeatother

\makeatletter
\let\old@makecaption=\@makecaption
\def\@makecaption{\small\old@makecaption}
\makeatother

\let\oldPhi=\Phi
\let\oldPsi=\Psi
\let\oldGamma=\Gamma
\let\oldDelta=\Delta
\let\oldSigma=\Sigma
\let\oldTheta=\Theta
\let\oldPi=\Pi
\renewcommand{\Phi}{\mathnormal{\oldPhi}}
\renewcommand{\Psi}{\mathnormal{\oldPsi}}
\renewcommand{\Gamma}{\mathnormal{\oldGamma}}
\renewcommand{\Sigma}{\mathnormal{\oldSigma}}
\renewcommand{\Delta}{\mathnormal{\oldDelta}}
\renewcommand{\Theta}{\mathnormal{\oldTheta}}
\renewcommand{\Pi}{\mathnormal{\oldPi}}

\newcommand{\gen}[1]{\mathfrak{#1}}

\newcommand{\smat}{\mathcal{S}}

\newcommand{\superN}{\mathcal{N}}
\newcommand{\fldZ}{\mathcal{Z}}
\newcommand{\fldX}{\mathcal{X}}

\newcommand{\gym}{g\indups{YM}}

\newcommand{\order}[1]{\mathcal{O}(#1)}

\newcommand{\Reals}{\mathbb{R}}

\newcommand{\MMM}[2]{{\arraycolsep0pt\begin{array}[b]{c}\makebox[0cm]{$\atopfrac{#2}{\downarrow}$}\\#1\end{array}}}

\ifx\genfrac\sdflkaj
\newcommand{\atopfrac}[2]{{{#1}\above0pt{#2}}}
\else
\newcommand{\atopfrac}[2]{\genfrac{}{}{0pt}{}{#1}{#2}}
\fi
\newcommand{\sfrac}[2]{{\textstyle\frac{#1}{#2}}}
\newcommand{\half}{\sfrac{1}{2}}
\newcommand{\ihalf}{\sfrac{i}{2}}

\newcommand{\indups}[1]{_{\mathrm{\scriptscriptstyle #1}}}

\newcommand{\rep}[1]{{\mathbf{#1}}}

\newcommand{\lvl}[1]{^{\mathrm{#1}}}

\newcommand{\lrbrk}[1]{\left(#1\right)}
\newcommand{\bigbrk}[1]{\bigl(#1\bigr)}

\newcommand{\comm}[2]{[#1,#2]}

\newcommand{\acomm}[2]{\{#1,#2\}}

\newcommand{\bigabs}[1]{\bigl|#1\bigr|}
\newcommand{\bigeval}[1]{#1\big|}

\newcommand{\set}[1]{\{#1\}}
\newcommand{\state}[1]{\mathopen{|}#1\mathclose{\rangle}}

\newcommand{\alg}[1]{\mathfrak{#1}}
\newcommand{\grp}[1]{\mathrm{#1}}

\newcommand{\nn}{\nonumber}
\newcommand{\nln}{\nonumber\\}
\newcommand{\nl}[1][0pt]{\nonumber\\[#1]&\hspace{-4\arraycolsep}&\mathord{}}
\newcommand{\nlnum}{\\&\hspace{-4\arraycolsep}&\mathord{}}
\newcommand{\earel}[1]{\mathrel{}&\hspace{-2\arraycolsep}#1\hspace{-2\arraycolsep}&\mathrel{}}
\newcommand{\eq}{\earel{=}}

\def\[{\begin{equation}}
\def\]{\end{equation}}
\def\<{\begin{eqnarray}}
\def\>{\end{eqnarray}}

\makeatletter
\def\mr@ignsp#1 {\ifx\:#1\@empty\else #1\expandafter\mr@ignsp\fi}%
\newcommand{\multiref}[1]{\begingroup
\xdef\mr@no@sparg{\expandafter\mr@ignsp#1 \: }%
\def\mr@comma{}%
\@for\mr@refs:=\mr@no@sparg\do{\mr@comma\def\mr@comma{,}\ref{\mr@refs}}%
\endgroup}
\makeatother

\newcommand{\hypref}[2]{\ifx\href\asklfhas #2\else\href{#1}{#2}\fi}

\newcommand{\secref}[1]{Sec.~\multiref{#1}}

\newcommand{\appref}[1]{App.~\multiref{#1}}

\newcommand{\tabref}[1]{Tab.~\multiref{#1}}

\renewcommand{\eqref}[1]{(\multiref{#1})}

\ifx\href\asklfhas\newcommand{\href}[2]{#2}\fi
\newcommand{\arxivno}[1]{\href{http://arxiv.org/abs/#1}{#1}}

\begin{document}

\thispagestyle{empty}
\begin{flushright}\footnotesize
\texttt{\arxivno{hep-th/0511082}}\\
\texttt{PUTP-2181}\\
\texttt{NSF-KITP-05-92}\\
\vspace{0.5cm}
\end{flushright}
\vspace{0.5cm}

\renewcommand{\thefootnote}{\fnsymbol{footnote}}
\setcounter{footnote}{0}

\begin{center}
{\Large\textbf{\mathversion{bold}
The $\alg{su}(2|2)$ Dynamic S-Matrix
}\par} \vspace{1cm}

\textsc{Niklas Beisert} \vspace{5mm}

\textit{Joseph Henry Laboratories\\
Princeton University\\
Princeton, NJ 08544, USA} \vspace{3mm}

\texttt{nbeisert@princeton.edu}\\
\par\vspace{1cm}

\vfill

\textbf{Abstract}\vspace{5mm}

\begin{minipage}{12.7cm}
We derive and investigate the S-matrix 
for the $\alg{su}(2|3)$ dynamic spin chain 
and for planar $\superN=4$ super Yang-Mills. 
Due to the large amount of residual symmetry in the excitation picture,
the S-matrix turns out to be fully constrained up to an overall phase. 
We carry on by diagonalising it and obtain
Bethe equations for periodic states.
This proves an earlier proposal for the asymptotic Bethe 
equations for the $\alg{su}(2|3)$ dynamic spin chain and for $\superN=4$ SYM.
\end{minipage}

\vspace*{\fill}

\end{center}

\newpage
\setcounter{page}{1}
\renewcommand{\thefootnote}{\arabic{footnote}}
\setcounter{footnote}{0}

\section{Introduction and Conclusions}

In general, computations in perturbative field theories 
are notoriously intricate.
Recently, the discovery and application of integrable structures 
in planar four-dimensional gauge theories, 
primarily in conformal $\superN=4$ super Yang Mills theory,
has lead to drastic simplifications in determining some quantities.
In particular, planar anomalous dimensions of local operators
can be mapped to energies of quantum spin chain states 
thus establishing some relation to topics of condensed matter physics.
The Hamiltonian of this system is completely integrable 
at one loop \cite{Lipatov:1997vu,Minahan:2002ve,Beisert:2003yb}
and apparently even at higher loops \cite{Beisert:2003tq,Beisert:2003ys},
cf.~the reviews \cite{Beisert:2004ry,Beisert:2004yq,Zarembo:2004hp,Plefka:2005bk}.
This remarkable feature shows promise that the planar spectrum might be
described \emph{exactly} by some sort of Bethe equation. 
Bethe equations at the one-loop level were given in \cite{Minahan:2002ve,Beisert:2003yb}.
At higher loops some similarity 
of the exact gauge theory result 
\cite{Beisert:2003tq,Beisert:2003ys,Eden:2004ua}
with the Inozemtsev spin chain 
\cite{Inozemtsev:1989yq,Inozemtsev:2002vb} can be observed 
and Bethe equations for the $\alg{su}(2)$ sector up to 
three loops were found in \cite{Serban:2004jf}.
They were then generalised to the other two rank-one sectors,
$\alg{su}(1|1)$ and $\alg{sl}(2)$, in \cite{Staudacher:2004tk}.
All-loop asymptotic Bethe equations for the $\alg{su}(2)$ sector 
with some more desirable features for $\superN=4$ SYM were proposed in \cite{Beisert:2004hm}.
Putting together all the above pieces of a puzzle, 
asymptotic Bethe equations for the complete model were finally proposed 
in \cite{Beisert:2005fw}.

Bethe equations have since proved very fruitful for the study
of the AdS/CFT correspondence
\cite{Maldacena:1998re,Gubser:1998bc,Witten:1998qj}
and certain limits of it involving large spins 
\cite{Berenstein:2002jq,Gubser:2002tv,Frolov:2002av}.
On the string theory side of the correspondence
integrability has been established for the classical theory in \cite{Mandal:2002fs,Bena:2003wd}
and evidence for quantum integrability exists 
\cite{Arutyunov:2004vx,Berkovits:2004xu}.
The results for spinning strings \cite{Frolov:2003qc,Beisert:2003xu,Frolov:2003xy,Beisert:2003ea}
and near plane wave strings \cite{Callan:2003xr,Callan:2004uv}
have lead to new insights into the correspondence,
see the reviews \cite{Beisert:2004ry,Beisert:2004yq,Zarembo:2004hp,Plefka:2005bk,Tseytlin:2003ii,Swanson:2005wz}
for details and further references.
\medskip

The Bethe equations for $\superN=4$ SYM mentioned above
have many desired features and they seem to work, 
but it is fair to say that their origin remains obscure.
At the one-loop level the Hamiltonian involves nearest-neighbour
interactions only. One can therefore resort to the well-known 
R-matrix formalism to derive and study the Bethe equations. 
At higher loops the interactions of the Hamiltonian become more complex: 
Their range increases with the loop order \cite{Beisert:2003tq}. 
Moreover, the length of the spin chain starts to fluctuate, 
sites are created or destroyed dynamically \cite{Beisert:2003ys}.
These types of spin chains have not been considered extensively and
there is no theoretical framework (yet); the higher-loop Bethe equations
are at best well-tested conjectures. 
The situation improved somewhat with the proposal of \cite{Staudacher:2004tk}.
By applying the asymptotic coordinate space Bethe ansatz
\cite{Bethe:1931hc,Sutherland:1978aa}, one may extract
a two-particle \emph{S-matrix} from the perturbative Hamiltonian. 
Assuming factorised scattering, this S-matrix is, like the R-matrix, 
a nearest-neighbour operator. 
At this stage one can therefore revert to the familiar framework. 
The resulting asymptotic Bethe equations turn out to reproduce the
spectrum accurately \cite{Staudacher:2004tk}.

The perturbative S-matrices for all three rank-one sectors,
$\alg{su}(2)$, $\alg{su}(1|1)$ and $\alg{sl}(2)$,  
were derived in \cite{Staudacher:2004tk} up to three loops.
The S-matrix in the $\alg{su}(2)$ sector coincides with the all-loop
conjecture of \cite{Beisert:2004hm} which can be read off directly from 
the asymptotic Bethe equations.
Corresponding all-loop conjectures for the other two
rank-one sectors were set up in \cite{Beisert:2005fw};
they have a similarly concise form.
All these rank-one sectors can be joined into one
larger sector with $\alg{su}(1,1|2)$ symmetry for which 
an all-loop S-matrix was also conjectured.
This conjecture agrees with the Hamiltonian derived in \cite{Beisert:2003ys}
up to three loops in the subsector where both results apply.
\medskip

It is the purpose of the present investigation to find the complete 
S-matrix for planar $\superN=4$ SYM. 
This will allow to put the asymptotic Bethe equations 
conjectured in \cite{Beisert:2005fw} on a solid footing
and hopefully give us a better understanding of
the asymptotic Bethe ansatz as well as 
the integrable structures in gauge theory in general.
The partial results mentioned above as well as the 
resulting Bethe equations suggest that 
also the complete S-matrix might have a simple form
valid to all perturbative orders. 
A major problem that one has to deal with in finding the S-matrix
is that the complete spin chain is dynamic \cite{Beisert:2003ys},
its length fluctuates. 
In the excitation picture this might appear
not to be a problem as the number of excitations is preserved, 
but even there one finds flavour fluctuations 
which may appear problematic \cite{Beisert:2005fw}.

An important property of S-matrices is their symmetry. 
Often they can be constructed from symmetry considerations 
and a few additional properties. 
Also the S-matrices appearing in sectors of planar $\superN=4$ SYM
are largely constrained by their symmetry. A somewhat unusual
feature of these particular S-matrices is that the representations 
in which the excitations transform obey a dispersion relation \cite{Beisert:2005wm}. 
This can be related to the fact that the 
Hamiltonian is part of the symmetry algebra and not some central
generator as for most spin chain models.
For instance, in the $\alg{su}(1|2)$ sector the all-loop
form of the S-matrix has manifest $\alg{su}(1|1)$ symmetry. 
The full symmetry algebra of $\superN=4$ SYM is $\alg{psu}(2,2|4)$.
The S-matrix in the excitation picture, however, is manifestly invariant
only under a residual algebra which preserves the excitation number.
In this case the residual algebra is $\alg{psu}(2|2)^2\ltimes \Reals$,
cf.~\cite{Beisert:2004ry}.
The excitations transform in a $\rep{(2|2)}$ representation
under each $\alg{psu}(2|2)$ factor.
Both factors share a common central charge
$\gen{C}$ which takes the role of the Hamiltonian.
To be precise, we will introduce two further unphysical central charges 
related to the dynamic nature of the spin chain.

For the construction of the S-matrix
it turns out to be very helpful that 
the algebra splits into two (equal) parts:
The complete S-matrix can be constructed as
a product of two S-matrices, each transforming only
under one of the subalgebras.
Moreover, as the particle representations
of both subalgebras are isomorphic,
it is sufficient to construct only one S-matrix
with $4^4$ components instead of $(4^2)^4$.
We can therefore work with a reduced set of $(2|2)$ excitations
and an S-matrix transforming under the reduced algebra $\alg{su}(2|2)$. 
Incidentally, this coincides with the S-matrix of the 
maximally compact sector of $\superN=4$ SYM 
which is the $\alg{su}(2|3)$ dynamic spin chain 
investigated in \cite{Beisert:2003ys}.
\medskip

As a first step towards the S-matrix, 
we investigate the residual algebra in \secref{sec:Alg}
and find a suitable representation for the excitations. 
On the one hand, the representation $\rep{(2|2)}$ is almost 
the fundamental of $\alg{su}(2|2)$, but it requires
a trivial central charge $C=\pm \half$. 
On the other hand, the central charge of
$\alg{su}(2|2)$ represents the energy and we know that it
is not quantised in units of $\half$. 
To circumvent this seeming paradox 
we enlarge the algebra by two central charges $\gen{P,K}$.%
\footnote{The letters $\mathfrak{J,R,L,Q,S,C,P,K}$ of the \texttt{$\mathtt{\backslash}$mathfrak} alphabet
correspond to $\mathrm{J,R,L,Q,S,C,P,K}$.}
This is indeed possible and allows for a non-trivial 
$\rep{(2|2)}$ representation 
with one free continuous degree of freedom.
We construct this representation subsequently. 
The two additional central charges 
can be related to gauge transformations
which act non-trivially on individual fields;
nevertheless they must annihilate 
gauge invariant combinations of fields
and therefore we can return to $\alg{su}(2|2)$ 
as the global symmetry.

Having understood the representation of the symmetry algebra,
we construct the S-matrix as an invariant permutation operator 
on two-excitation states in \secref{sec:Smat}. 
Astonishingly, the S-matrix is uniquely determined up to an overall phase.
This fact may be attributed to the uniqueness of $\superN=4$ SYM. 
An unconstrained overall phase 
is a common problem of constructive methods. 
In fact, the model in \cite{Beisert:2003ys}
leaves some degrees of freedom which are 
reflected by this phase \cite{Beisert:2005fw}.
We then study the properties of the
S-matrix and find that it naturally satisfies the
Yang-Baxter equation. This is a necessary condition
for factorised scattering and integrability. 
Assuming that integrability holds, we outline the construction
of eigenstates of the Hamiltonian. 

In \secref{sec:Diag} we perform the nested Bethe ansatz \cite{Yang:1967bm} 
on this S-matrix. 
This leads to a completely diagonalised S-matrix which can be
employed for the asymptotic Bethe equations. 
We then study the symmetry properties of the 
equations and the remaining phase.
It is also straightforward to ``square{}'' the 
S-matrix and obtain Bethe equations for
$\superN=4$ SYM, cf.~\secref{sec:N4}.
We can thus prove the validity of the conjectured
asymptotic Bethe equations of \cite{Beisert:2005fw}
(up to the unknown abelian phase and under the
assumption of integrability).
Among other things, 
this represents a further piece of evidence for the correctness
of the conjecture for the three-loop planar anomalous dimensions
of twist-two operators \cite{Kotikov:2004er}.
The conjecture was based on an explicit three-loop
computation in QCD \cite{Moch:2004pa,Vogt:2004mw}
and a lift to $\superN=4$ SYM by means of ``transcendentality{}'' counting.
They were subsequently reproduced in the
asymptotic Bethe ansatz for the $\alg{sl}(2)$ sector \cite{Staudacher:2004tk}.
The derivation of the latter required a 
relation to hold between the S-matrices of
rank-one sectors; here we can identify the group theoretical origin
of this relation.
\medskip

The only missing piece of information for the complete S-matrix
is its abelian phase. 
Its determination is prevented here because 
it is neither constrained by representation theory nor by the Yang Baxter relation.
A frequently employed constraint in two-dimensional integrable sigma models,
see e.g.~\cite{Zamolodchikov:1978xm}, 
is a crossing relation for the S-matrix whose existence remains obscure here. 
Furthermore, the pole structure of the S-matrix might lead to some constraints. 
The results in \appref{sec:Singlet} concerning a curious singlet state 
represent some (failed) attempts in this direction; 
it is not (yet) clear how to make sense of them.

A possible direction for future research is to perform
a similar investigation for the S-matrix of
the IIB string theory on $AdS_5\times S^5$.
The classical diagonalised S-matrix elements 
can be read off from an integral representation 
of the classical spectral curve in \cite{Kazakov:2004qf,Kazakov:2004nh,Beisert:2004ag,Schafer-Nameki:2004ik,Beisert:2005bm}
and the proposed Bethe equations for quantum strings
\cite{Arutyunov:2004vx,Beisert:2005fw}.
Clearly, the actual non-diagonalised S-matrix is important
as the underlying structure of the Bethe ansatz,
cf.~\cite{Das:2004hy,Arutyunov:2004yx,Alday:2005gi,Das:2005hp,Alday:2005jm,Arutyunov:2005hd} 
for some results in this direction.
Due to the AdS/CFT correspondence, one might expect the 
S-matrix to have the same or at least a very similar form
and an explicit derivation would be very valuable.
Unless there are more powerful constraints here, 
we should again expect an undetermined phase.
The phase can be determined perturbatively by comparison
to spinning string states, cf.~\cite{Gubser:2002tv,Frolov:2002av}
and the reviews \cite{Tseytlin:2003ii,Plefka:2005bk}.
Some leading quantum corrections 
to these states and methods to deal with them 
in the Bethe ansatz have recently become available 
\cite{Frolov:2003tu,Frolov:2004bh,Park:2005ji,
Beisert:2005mq,Hernandez:2005nf,Beisert:2005bv,Schafer-Nameki:2005tn,
Beisert:2005cw,Schafer-Nameki:2005is,Gromov:2005gp}.
A somewhat different approach (for a somewhat different model) 
might also lead to Bethe equations for quantum
strings \cite{Mann:2004jr,Mann:2005ab}.

Another possible application for the current results
is plane wave matrix theory 
\cite{Berenstein:2002jq,Lin:2005nh}.
This theory leads to a very similar spin chain model
\cite{Kim:2003rz,Fischbacher:2004iu,Klose:2005aa},
which is however not completely integrable beyond leading order \cite{Beisert:2005wv}.
Nevertheless, it has an $\alg{su}(3|2)$ sector and the present result
about the two-particle S-matrix certainly does apply.
This S-matrix satisfies the YBE, factorised scattering is thus self-consistent.
The important question however is 
whether the multi-particle S-matrix does indeed factorise;
is the $\alg{su}(3|2)$ sector of PWMT integrable?

Finally, we should point out that the current analyses are
justified only in the asymptotic region:
At high orders in perturbation theory 
there are interactions whose range may 
exceed the length of a spin chain state, 
the so-called wrapping interactions
\cite{Beisert:2003ys,Serban:2004jf,Beisert:2004hm}.
The asymptotic Bethe equations should only be trusted up to 
to this perturbative order which depends on the length of the chain
(which itself if somewhat ill-defined in dynamic chains). 
Unfortunately, it is very hard to make precise statements because
wrapping interactions are practically inaccessible by constructive methods 
of the planar Hamiltonian
(and four-loop field theory computations are somewhat beyond our
current possibilities). 
Nevertheless when considering
the finite-$N$ algebra, they should be incorporated naturally \cite{Zwiebel:2005er}.
It is very likely that the asymptotic Bethe equations receive
corrections, either in the form of corrections to the 
undetermined phase in an effective field theory sense or, 
preferably, by improved equations. The thermodynamic
Bethe ansatz may provide a suitable framework here \cite{Ambjorn:2005wa}.

\section{The Asymptotic $\alg{su}(2|2)$ Algebra}
\label{sec:Alg}

In the following we introduce the spin chain model. We then
consider asymptotic states of an infinitely long spin chain
and investigate the residual symmetry which preserves the number of excitations 
of states.

\subsection{The $\alg{su}(2|3)$ Dynamic Spin Chain Model}

In \cite{Beisert:2003ys} a spin chain with 
$\alg{su}(2|3)$ symmetry and fundamental matter was considered.
This spin chain arises as a sector of 
perturbative $\grp{U}(N)$ $\superN=4$ super Yang-Mills theory in the large-$N$ limit.
The spin $\fldX$ at each site can take one out of five orientations 
$\fldX\in\set{\fldZ,\phi^1,\phi^2\mathpunct{|}\psi^1,\psi^2}$.
The first three are bosonic states, the remaining two are fermions;
in a $\superN=1$ notation they represent the three scalar 
fields and the two spin orientations of the gluino.
A generic state $\state{\Psi}$ is a linear combination of 
basic states, e.g.
\[
\state{\Psi}=
\ast\state{\fldZ\phi^1\fldZ\fldZ\psi^2\fldZ\ldots\phi^1}+
\ast\state{\psi^1\phi^2\fldZ\fldZ\psi^2\ldots\fldZ}+\ldots\,.
\]
Such a state represents a single-trace gauge invariant local operator.
The spin chain is closed and physical states are cyclic, 
they must be invariant under cyclic 
permutations of the spin sites taking into account
the statistics of the fields.
This corresponds to cyclicity of the trace in gauge theory.

The states transform under a symmetry algebra $\alg{su}(2|3)$
which is a subalgebra of the superconformal algebra $\alg{psu}(2,2|4)$
of $\superN=4$ SYM.
The $\alg{gl}(1)$ generator of this algebra is associated 
with the energy, we shall call it the Hamiltonian;
in $\superN=4$ SYM it is related to the dilatation generator.
Thus, finding the spectrum of this operator is physically interesting,
it contains the planar anomalous dimensions of the
local operators in the $\alg{su}(2|3)$ sector.
A family of representations of $\alg{su}(2|3)$
on spin chain states
was constructed in \cite{Beisert:2003ys}.
The family was parametrised by the coupling constant $g$
related to the 't~Hooft coupling constant by
\[
g^2=\frac{\lambda}{8\pi^2}=\frac{\gym^2N}{8\pi^2}\,.
\]
At $g=0$ the representation is merely the tensor product
of fundamental representations. The deformations around
this point can be constructed in perturbation theory. 
This was done in \cite{Beisert:2003ys} 
up to fourth order for all generators
and up to sixth order for the Hamiltonian.
The constraining property of the representation was that the 
generators must act locally on the spin chain with 
a maximum range determined by the order in $g$. 
At a finite value of $g$, the action is therefore
\emph{long-ranged}.
The action is also \emph{dynamic}, 
the generators are allowed to change the number
spin chain sites $L$: the length $L$ fluctuates.

\subsection{Asymptotic States}

Let us define a vacuum state composed from only $\fldZ$'s. 
We shall start with an infinitely long vacuum
\[\label{eq:Alg.Vac}
\state{0}\lvl{I}=\state{\ldots \fldZ\fldZ\ldots\fldZ\fldZ\ldots}.
\]
In fact, physical states have a finite length and are periodically
identified.
As pointed out in \cite{Staudacher:2004tk},
it is however sufficient to consider periodic states 
on an infinite chain to obtain the correct spectrum
up to a certain accuracy. This is what will be called the \emph{asymptotic regime}. 
We might then consider a generic \emph{asymptotic state} as an excitation 
of the vacuum, such as
\[\label{eq:Alg.State}
\state{\fldX^{}_1\ldots \fldX''_K}\lvl{I}=
\sum_{n_1\ll\ldots\ll n_K} e^{ip_1n_1}\ldots e^{ip_Kn_K}\, 
\state{\ldots\fldZ\fldZ\ldots\MMM{\fldX}{n_1}\ldots\MMM{\fldX'}{\ldots}\ldots\MMM{\fldX''}{n_K}\ldots\fldZ\fldZ\ldots}.
\]
The superscript ``I'' of the state implies that we have 
screened out all vacuum fields $\fldZ$. Here ``I'' refers
to the first level of screening; later, at higher levels, 
more fields will be screened.
The excitations $\fldX\in\set{\phi^1,\phi^2\mathpunct{|}\psi^1,\psi^2}$
have the same order with which they appear in the original spin chain. 
The subscript $k=1,\ldots, K$ of an excitation indicates that 
$\fldX_k$ carries a definite momentum $p_k$ along the original spin chain.

In \eqref{eq:Alg.State} we have assumed that the
excitations are well-separated, $n_{k}\ll n_{k+1}$,
so that the range of interactions is always 
smaller than the minimum separation. 
Then the interactions act on only one excitation 
at a time which is a major simplification;
this is our notion of asymptotic states. 
Of course also the states with nearby excitations 
are important, but for the
determination of asymptotic eigenstates and energies 
their contribution can be summarised by the S-matrix which will be considered
in the next \secref{sec:Smat}.

\subsection{The Algebra}
\label{sec:Alg.Alg}

The spin chain states transform under the full symmetry algebra $\alg{su}(2|3)$
and so do the asymptotic states. However, the number of excitations, $K$,
is not preserved. It is only preserved by a subalgebra of $\alg{su}(2|3)$,
namely $\alg{su}(2|2)$, let us therefore restrict to it. 
This algebra $\alg{su}(2|2)$ consists of 
the $\alg{su}(2)\times\alg{su}(2)$ rotation generators 
$\gen{R}^a{}_b$, $\gen{L}^\alpha{}_\beta$, the 
supersymmetry generators 
$\gen{Q}^\alpha{}_b$, $\gen{S}^a{}_\beta$ 
and the central charge $\gen{C}$.
The non-trivial commutators are
\<\label{eq:Alg.CommSU22}
\comm{\gen{R}^a{}_b}{\gen{J}^c}\eq\delta^c_b\gen{J}^a
  -\half \delta^a_b\gen{J}^c,
\nln
\comm{\gen{L}^\alpha{}_\beta}{\gen{J}^\gamma}\eq
  \delta^\gamma_\beta\gen{J}^\alpha
  -\half \delta^\alpha_\beta\gen{J}^\gamma,
\nln
\acomm{\gen{Q}^\alpha{}_a}{\gen{S}^b{}_\beta}
\eq 
\delta^b_a\gen{L}^\alpha{}_\beta
+\delta^\alpha_\beta\gen{R}^b{}_a
+\delta^b_a\delta^\alpha_\beta\gen{C},
\>
where $\gen{J}$ represents any generator with the appropriate index.
For later convenience we enlarge the algebra by two central charges%
\footnote{Central extensions of Lie superalgebras were investigated
in \cite{Iohara:2001aa}. 
The $\alg{psu}(2|2)$ algebra constitutes a special case with up to three central extensions. 
I thank F.~Spill for pointing out this reference to me.}
$\gen{P},\gen{K}$ to $\alg{su}(2|2)\ltimes \Reals^2$
\<\label{eq:Alg.CommExt}
\acomm{\gen{Q}^\alpha{}_a}{\gen{Q}^\beta{}_b}
\eq 
\varepsilon^{\alpha\beta}\varepsilon_{ab}\gen{P},\nln
\acomm{\gen{S}^a{}_\alpha}{\gen{S}^b{}_\beta}
\eq 
\varepsilon^{ab}\varepsilon_{\alpha\beta}\gen{K}.
\>
These shall have zero eigenvalue on physical states and thus
the algebra on physical states is effectively $\alg{su}(2|2)$.
The extension is necessary because the representations
of $\alg{su}(2,2)$ are too restrictive for the
excitation picture. 

The enlarged algebra $\alg{psu}(2|2)\ltimes \Reals^3$ 
is a contraction of the exceptional superalgebra $\alg{d}(2,1;\epsilon,\Reals)$ 
with $\epsilon\to 0$. 
The triplet of central charges $\gen{P}$, $\gen{K}$ and $\gen{C}$ is the contraction
of the $\alg{sp}(2,\Reals)$ factor while the rotation generators 
$\gen{R},\gen{L}$ form the $\alg{so}(4)=\alg{su}(2)^2$ part.
See \appref{sec:contract} for details of this construction.

\subsection{The Representation}
\label{sec:Alg.Rep}

Let us represent $\alg{su}(2|2)$ on a $2|2$-dimensional space.
We label the states by $\state{\phi^a}\lvl{I}$ and $\state{\psi^\alpha}\lvl{I}$.
These should be considered single excitations \eqref{eq:Alg.State}
of the level-I vacuum $\state{0}\lvl{I}$ in \eqref{eq:Alg.Vac}.
Each $\alg{su}(2)$ factor should act canonically on either of the 
two-dimensional subspaces%
\footnote{The $\alg{su}(2)$ algebra generates a compact group
whose unitary/finite-dimensional representations cannot be deformed continuously.}
\<\label{eq:Rep.ActionRL}
\gen{R}^a{}_b\state{\phi^c}\lvl{I}\eq\delta^c_b\state{\phi^a}\lvl{I}
  -\half \delta^a_b\state{\phi^c}\lvl{I},
\nln
\gen{L}^\alpha{}_\beta\state{\psi^\gamma}\lvl{I}\eq\delta^\gamma_\beta\state{\psi^\alpha}\lvl{I}
  -\half \delta^\alpha_\beta\state{\psi^\gamma}\lvl{I}.
\>
The supersymmetry generators should also act in a manifestly
$\alg{su}(2)\times\alg{su}(2)$ covariant way.
The most general transformation rules are thus
\<\label{eq:Rep.ActionQS}
\gen{Q}^\alpha{}_a\state{\phi^b}\lvl{I}\eq a\,\delta^b_a\state{\psi^\alpha}\lvl{I},\nln
\gen{Q}^\alpha{}_a\state{\psi^\beta}\lvl{I}\eq b\,\varepsilon^{\alpha\beta}\varepsilon_{ab}\state{\phi^b\fldZ^+}\lvl{I},\nln
\gen{S}^a{}_\alpha\state{\phi^b}\lvl{I}\eq c\,\varepsilon^{ab}\varepsilon_{\alpha\beta}\state{\psi^\beta\fldZ^-}\lvl{I},\nln
\gen{S}^a{}_\alpha\state{\psi^\beta}\lvl{I}\eq d\,\delta^\beta_\alpha\state{\phi^a}\lvl{I}.
\>
For the moment we shall ignore the symbols
$\fldZ^\pm$ inserted into the states.
We find that the closure of $\acomm{\gen{Q}}{\gen{S}}=\ldots$ 
\eqref{eq:Alg.CommSU22,eq:Alg.CommExt} requires
$ad-bc=1$. The central charge is then given by 
\[\label{eq:Rep.ActionC}
\gen{C}\state{\fldX}\lvl{I}=C\state{\fldX}\lvl{I}=\half (ad+bc)\,\state{\fldX}\lvl{I}
\]
where $\state{\fldX}\lvl{I}$ is any of the states $\state{\phi^a}\lvl{I}$
or $\state{\psi^\alpha}\lvl{I}$.
For $\alg{su}(2|2)$ we should furthermore impose 
$\acomm{\gen{Q}}{\gen{Q}}=\acomm{\gen{S}}{\gen{S}}=0$
which fixes $ab=0$ and $cd=0$.
The two solutions to these equations lead to a central charge $C=\pm \half$
and correspond to the fundamental representations of $\alg{su}(2|2)$.
This would lead to the model introduced in \cite{Essler:1992aa}
which is the correct description of gauge theory at leading order, 
but not at higher loops.

In order to find more interesting solutions with non-trivial central charge
we relax the condition $\acomm{\gen{Q}}{\gen{Q}}=\acomm{\gen{S}}{\gen{S}}=0$
and allow for non-trivial central charges $\gen{P},\gen{K}$.
Closure of the symmetry algebra requires the action of the additional generators 
to be
\<\label{eq:Rep.ActionPK}
\gen{P}\state{\fldX}\eq ab\,\state{\fldX\fldZ^+},\nln
\gen{K}\state{\fldX}\eq cd\,\state{\fldX\fldZ^-}.
\>
Of course we are interested in representations of the original
$\alg{su}(2|2)$ algebra and not of some enlarged one. 
Therefore we are bound to constrain 
the action of $\gen{P}$ and $\gen{K}$ to zero. For the above 
representation we are back at where we started and 
there is only the fundamental representation. 
The improvement of this point of view comes about when 
we consider tensor products. 
Then, only the action of the \emph{overall} generators 
$\gen{P}$ and $\gen{K}$ must be zero leaving some
degrees of freedom among the individual representations.

\subsection{Dynamic Spin Chains}
\label{sec:Alg.Dyn}

To match the representation to excitations of
the dynamic $\alg{su}(2|3)$ spin chain \cite{Beisert:2003ys},
we note that $\fldZ^+$ should be considered as the insertion 
of a field $\fldZ$ into the original chain;
likewise $\fldZ^-$ removes a field. 
Let us consider an excitation with a definite momentum 
on an infinite spin chain 
\[
\state{\fldX}\lvl{I}=\sum\nolimits_n e^{ipn}\, 
\state{\ldots\fldZ\fldZ\ldots\MMM{\fldX}{n}\ldots\fldZ\fldZ\ldots}.
\]
When we insert or remove a background field $\fldZ$ in front of the excitation 
we obtain
\[
\state{\fldZ^\pm\fldX}\lvl{I}
= \sum\nolimits_n e^{ipn} \,
\state{\ldots\fldZ\fldZ\ldots\MMM{\fldX}{n\pm 1}\ldots\fldZ\fldZ\ldots}
= \sum\nolimits_n e^{ipn\mp ip}\, 
\state{\ldots\fldZ\fldZ\ldots\MMM{\fldX}{n}\ldots\fldZ\fldZ\ldots},
\]
i.e.~we can always shift the operation $\fldZ^\pm$ to the very right of the 
asymptotic state
and pick up factors of $\exp(\mp ip)$
\[\label{eq:Alg.Shift}
\state{\fldZ^\pm\fldX}\lvl{I}=
e^{\mp ip}\,\state{\fldX\fldZ^\pm}\lvl{I}.
\]

The action of $\gen{P}$ on a tensor product gives
\[
\gen{P}\state{\fldX_1\ldots \fldX_K}\lvl{I}
=
P
\state{\fldX_1\ldots \fldX_K\fldZ^+}\lvl{I},
\qquad
P=\sum_{k=1}^K
a_kb_k\prod_{l=k+1}^K e^{-ip_l}
\]
and should vanish on physical states. Physical states are thus defined
by the condition that the central charge $P$ vanishes. On the other hand 
we know that physical states are cyclic, they have zero total momentum.
Indeed $P=0$ coincides with the zero momentum condition
provided that we set $a_kb_k=\alpha (e^{-ip_k}-1)$. 
Then the sum telescopes and becomes
\[
P=
\alpha\sum_{k=1}^K
(e^{-ip_k}-1)\prod_{l=k+1}^K e^{-ip_l}
=
\alpha\lrbrk{\prod_{k=1}^K e^{-ip_k}-1}.
\]
The first term is the eigenvalue of the right shift operator.
When we set $c_kd_k=\beta (e^{ip_k}-1)$ we obtain the same constraint 
from a vanishing action of $\gen{K}$ 
\[
\gen{K}\state{\fldX_1\ldots \fldX_K}\lvl{I}
=
\beta\lrbrk{
\prod_{k=1}^K e^{ip_k}-1}
\state{\fldX_1\ldots \fldX_K\fldZ^-}\lvl{I}.
\]

We can now write the action of $\gen{P},\gen{K}$ in \eqref{eq:Rep.ActionPK} as 
\<
\gen{P}\state{\fldX}\lvl{I}\eq \alpha\state{\fldZ^+\fldX}\lvl{I}-\alpha\,\state{\fldX\fldZ^+}\lvl{I},\nln
\gen{K}\state{\fldX}\lvl{I}\eq \beta\state{\fldZ^-\fldX}\lvl{I}-\beta\,\state{\fldX\fldZ^-}\lvl{I}.
\>
Note that this reveals their nature as a
gauge transformation, $\gen{P}$ generates
the transformation
$\Psi\mapsto \alpha\comm{\fldZ}{\Psi}$.
Similarly, $\gen{K}$ generates
a somewhat unusual transformation 
$\Psi\mapsto \beta\comm{\fldZ^-}{\Psi}$,
which removes a field $\fldZ$.
Of course, physical states are gauge invariant 
and therefore should be annihilated by $\gen{P}$ and $\gen{K}$.

\subsection{Solution for the Coefficients}

Next we solve the central charge 
in terms of the momenta and obtain
\[\label{eq:Alg.Dispersion}
C=\sum_{k=1}^K C_k,\qquad C_k=\pm\half\sqrt{1+16\alpha\beta \sin^2(\half p_k)}\,.
\]
The central charge is the energy and consequently we have derived
the BMN-like energy formula \cite{Berenstein:2002jq} 
up to the value of the product $\alpha\beta$ which should play the role of 
the coupling constant.%
\footnote{This derivation of the energy formula should be similar to
the one in \cite{Santambrogio:2002sb}.}
To adjust to the correct coupling constant 
for $\superN=4$ SYM and the one used in \cite{Beisert:2003ys}
we set $\beta=g^2/2\alpha$.
We introduce new variables $x^+_k,x^-_k$ to replace the momenta $p_k$ and solve%
\footnote{The parameter $\gamma_k$ corresponds to a 
(momentum dependent) relative rescaling
of $\state{\phi^a}$ and $\state{\psi^\alpha}$
whereas $\alpha$ corresponds to a rescaling of $\fldZ$.}
\[\label{eq:Alg.abcd}
a_k=\gamma_k,\quad
b_k=-\frac{\alpha}{\gamma_k x_k^+}\,(x_k^+-x_k^-),\quad
c_k=\frac{ig^2\gamma_k}{2\alpha x_k^-}\,,\quad
d_k=-\frac{i}{\gamma_k }\,(x_k^+-x_k^-).
\]
For a hermitian representation we should choose
\[
|\gamma_k|=\bigabs{ix_k^--ix_k^+}{}^{1/2},\qquad 
|\alpha|=\sqrt{g^2/2}\,.
\]
The condition $a_kd_k-b_kc_k=1$ for the closure
of the algebra translates to 
\[\label{eq:Alg.xpmrel}
x_k^+
+\frac{g^2}{2x_k^+}
-x_k^-
-\frac{g^2}{2x_k^-}
=i.
\]
Finally, the momentum and central charge are given by 
\[\label{eq:Alg.MomEng}
e^{ip_k}=\frac{x^+_k}{x^-_k}\,,\qquad
C_k
=\frac{1}{2}+\frac{ig^2}{2x^+_k}-\frac{ig^2}{2x^-_k}
=-i x^+_k+ix^-_k-\frac{1}{2}\,.
\]
The physicality constraint 
$\gen{P}\state{\Psi}=\gen{K}\state{\Psi}=0$
is the zero-momentum condition
\[\label{eq:Alg.Phys}
1=\prod_{k=1}^K e^{ip_k}=\prod_{k=1}^K \frac{x^+_k}{x^-_k}\,.
\]

Interestingly, the dispersion relation \eqref{eq:Alg.Dispersion}
admits two solutions with a given momentum but opposite energies. 
This is a common feature of \emph{relativistic} quantum mechanics:
The two solutions can be interpreted as a regular particle
and a conjugate one propagating backwards in time. 
The conjugate excitation can be obtained from a regular one by
the substitution $x^\pm_k\mapsto -g^2/2x^\mp_k$
(or by $x^\pm_k\mapsto g^2/2x^\pm_k$ 
which inverts the momentum as well).

We might solve \eqref{eq:Alg.xpmrel} by \cite{Beisert:2004hm} 
\[\label{eq:Alg.xumap}
x_k^\pm=x(u_k\pm \ihalf),\qquad
x(u)=\half u+\half u\sqrt{1-2g^2/u^2}\,,\qquad
u(x)=x+\frac{g^2}{2x}\,.
\]
This may appear to yield only the positive energy solution,
it is however not possible to exclude the negative energy solution rigorously: 
In general $x^\pm$ are complex variables and 
the negative energy solution will always sneak in as the 
other branch of \eqref{eq:Alg.xumap}.
The branch cut may only be avoided in the 
\emph{non-relativistic} regime at $g\approx 0$,
where the perturbative gauge theory and the underlying spin chain
are to be found.

It seems that the appearance of conjugate excitations
is related to the puzzle observed in \cite{Minahan:2005jq}:
The $\alg{su}(2)$ sector of $\superN=4$ SYM does not
have a direct counterpart in string theory, 
but it is merely embedded in a larger $\alg{su}(2)\times \alg{su}(2)$
sector representing the isometry algebra of an $S^3$. 
This larger sector has excitations corresponding to a second $\alg{su}(2)$
which are related to the original ones by the map $x\mapsto g^2/2x$.
The reason why the conjugate excitations do not appear in 
gauge theory is related to perturbation theory.
They would have a non-vanishing anomalous dimension $-2$
at $g=0$ which is in conflict with the perturbative setup.
An interesting application of the conjugate excitations 
is presented in \appref{sec:Singlet} where
a peculiar composite of a regular excitation and its conjugate
is investigated.

\section{The S-Matrix}
\label{sec:Smat}

So far we have concentrated on asymptotic states (IR) 
and discarded the contributions from states
with nearby excitations (UV).
The latter become important when considering eigenstates
of the central charge $\gen{C}$ alias the Hamiltonian.
Luckily their inclusion can be summarised in the S-matrix
of the model.

\subsection{Sewing Eigenstates}

The symmetry algebra acts on 
the asymptotic states \eqref{eq:Alg.State}
as a tensor product representation:
All excitations are treated individually
and do not influence each other. 
This can however be true only in an asymptotic sense;
there are additional contributions from the boundaries
of the asymptotic regions where excitations come
too close. 
When interested in the exact
action of the algebra we must 
take these into account. 
This is achieved by sewing together the asymptotic regions 
in a way compatible with the algebra, e.g.
\[
\state{\Psi}=
a\state{\ldots \fldX^{}_k\fldX'_l\ldots}\lvl{I}+
b\state{\ldots (\fldX\fldX)_{kl}\ldots}\lvl{I}+
c\state{\ldots \fldX''_l\fldX'''_k\ldots}\lvl{I}\,.
\]
Here the left hand state is some asymptotic state,
the middle one represents contributions
with nearby excitations $k,l$ and in the 
right hand state the momenta of the excitations $k,l$ are interchanged.
There may be various linear combinations of different flavours
$\fldX,\fldX',\ldots$ which we do not specify here.
Clearly, the exact algebra transforms the coefficients $a,c$ independently
according to the asymptotic rules in \secref{sec:Alg.Rep}.
In addition, $b$ must be adjusted so that 
it yields the correct contributions to the boundaries
of the asymptotic regions. This relates $b$ to $a$ \emph{and} $b$ to $c$
and therefore $a$ \emph{with} $c$.
This means that asymptotic states can be completed to exact
states in a unique way compatible with the algebra. 
In particular, the coefficients of all asymptotic regions,
$a,c$ in the example, are related among each other.
As soon as this relation is known, it is no longer necessary to 
consider the non-asymptotic contributions. 
 
The completion of asymptotic states can be performed by the S-matrix.
The S-matrix $\smat\lvl{I}_{kl}$ is an operator
which interchanges two adjacent sites of the spin chain
at level I.
The affected sites are labelled by their momenta $p_k,p_l$
which are exchanged by $\smat\lvl{I}_{kl}$
\[
\smat\lvl{I}_{kl}\state{\ldots \fldX^{}_k\fldX'_l\ldots}\lvl{I}\mapsto 
\ast\state{\ldots \fldX''_l\fldX'''_k\ldots}\lvl{I}\,.
\]
The consistent completion of the above asymptotic state is then
\[
\state{\Psi}=
a\bigbrk{\state{\ldots \fldX^{}_k\fldX'_l\ldots}\lvl{I}+\mbox{non-asymp.}+
\smat\lvl{I}_{kl}\state{\ldots \fldX^{}_k\fldX'_l\ldots}\lvl{I}}.
\]
The requirement for asymptotic consistency is that the S-matrix 
commutes with the algebra, $\comm{\gen{J}_k+\gen{J}_l}{\smat\lvl{I}_{kl}}=0$,
where $\gen{J}_k$ is a generator of $\alg{su}(2|2)\ltimes \Reals^2$
acting on site $k$.

\subsection{Invariance}

Let us now construct the S-matrix
by acting on the state $\state{\fldX^{}_1\fldX'_2}$ with 
all possible combinations of spins $\fldX,\fldX'$.
We demand the exact invariance under $\alg{su}(2|2)\ltimes \Reals^2$
\[
\comm{\gen{J}_{1}+\gen{J}_{2}}{\smat\lvl{I}_{12}}=0.
\]
The commutators with the central charges $\gen{C},\gen{P},\gen{K}$
are automatically satisfied.
From commutators with the kinematic generators $\gen{R},\gen{L}$
the S-matrix takes a generic form determined by ten coefficient functions
$A_{12}=A(x_1,x_2)$ to $L_{12}$.
The form is presented at the top of \tabref{tab:SCoeff}. 
Remarkably, invariance under the dynamic generators $\gen{Q},\gen{S}$
leads to a \emph{unique} solution up to an undetermined overall function $S^0_{12}$.
To obtain the solution is straightforward but somewhat laborious;
we merely state the final result at the bottom of \tabref{tab:SCoeff}.

\begin{table}
\<
\smat\lvl{I}_{12}\state{\phi_1^a\phi_2^b}\lvl{I}
\eq
A_{12}\state{\phi_2^{\{a}\phi_1^{b\}}}\lvl{I}
+B_{12}\state{\phi_2^{[a}\phi_1^{b]}}\lvl{I}
+\half C_{12}\varepsilon^{ab}\varepsilon_{\alpha\beta}\state{\psi_2^\alpha\psi_1^\beta\fldZ^{-}}\lvl{I},
\nln
\smat\lvl{I}_{12}\state{\psi_1^\alpha\psi_2^\beta}\lvl{I}
\eq
D_{12}\state{\psi_2^{\{\alpha}\psi_1^{\beta\}}}\lvl{I}
+E_{12}\state{\psi_2^{[\alpha}\psi_1^{\beta]}}\lvl{I}
+\half F_{12}\varepsilon^{\alpha\beta}\varepsilon_{ab}\state{\phi_2^a\phi_1^b\fldZ^{+}}\lvl{I},
\nln
\smat\lvl{I}_{12}\state{\phi_1^a\psi_2^\beta}\lvl{I}
\eq
G_{12}\state{\psi_2^\beta\phi_1^{a}}\lvl{I}
+H_{12}\state{\phi_2^{a}\psi_1^\beta}\lvl{I},
\nln
\smat\lvl{I}_{12}\state{\psi_1^\alpha\phi_2^b}\lvl{I}
\eq
K_{12}\state{\psi_2^\alpha\phi_1^{b}}\lvl{I}
+L_{12}\state{\phi_2^{b}\psi_1^\alpha}\lvl{I}\nonumber.
\>

\<
A_{12}\eq S^0_{12}\,\frac{x_2^+-x_1^-}{x_2^--x_1^+}\,,\nln
B_{12}\eq S^0_{12}\,\frac{x_2^+-x_1^-}{x_2^--x_1^+}\lrbrk{1
             -2\,\frac{1-g^2/2x^-_2x^+_1}{1-g^2/2x^-_2x^-_1}\,
              \frac{x_2^+-x_1^+}{x_2^+-x_1^-}},\nln
C_{12}\eq 
S^0_{12}\,\frac{g^2\gamma_2\gamma_1}{\alpha x^-_2x^-_1}\,
  \frac{1}{1-g^2/2x^-_2x^-_1}\,\frac{x^+_2-x^+_1}{x_2^--x_1^+}\,,\nln
D_{12}\eq -S^0_{12},\nln
E_{12}\eq -S^0_{12}\,\lrbrk{1
         -2\,\frac{1-g^2/2x^+_2x^-_1}{1-g^2/2x^+_2x^+_1}\,
          \frac{x^-_2-x^-_1}{x^-_2-x^+_1}},\nln
F_{12}\eq -S^0_{12}\,\frac{2\alpha(x_2^+-x_2^-)(x_1^+-x_1^-)}{\gamma_2\gamma_1x^+_2x^+_1}\,
  \frac{1}{1-g^2/2x^+_2x^+_1}\,
  \frac{x^-_2-x^-_1}{x_2^--x_1^+}\,
  ,\nln
G_{12}\eq S^0_{12}\,\frac{x_2^+-x_1^+}{x_2^--x_1^+}\,,\nln
H_{12}\eq S^0_{12}\,\frac{\gamma_1}{\gamma_2}\,\frac{x_2^+-x_2^-}{x_2^--x_1^+}\,,\nln
K_{12}\eq S^0_{12}\,\frac{\gamma_2}{\gamma_1}\,\frac{x_1^+-x_1^-}{x_2^--x_1^+}\,,\nln
L_{12}\eq S^0_{12}\,\frac{x_2^--x_1^-}{x_2^--x_1^+}\,.\nonumber
\>
%
\caption{The dynamic $\alg{su}(2|2)$ S-matrix.}
\label{tab:SCoeff}
\end{table}

One may wonder why this S-matrix is uniquely determined.
It intertwines two modules and one should expect one
degree of freedom for each irreducible module in the tensor product.
Intriguingly, it appears that the tensor product 
is indeed irreducible. 
This may be the case because both factors are short (atypical). 
Their tensor product on the other hand
has $8|8$ components which is the smallest typical multiplet.
Note that the usual symmetrisations cannot be applied here, because 
both factors transform in distinct representations labelled by their
momenta $p_k$.
In verifying the invariance of the S-matrix, 
the following identities have proved useful
\<
\frac{x^+_1-x^+_2}{1-g^2/2x^-_1x^-_2}\eq
\frac{x^-_1-x^-_2}{1-g^2/2x^+_1x^+_2}\,,\nln
\frac{x^+_2-x^-_2-x^+_1+x^-_1}{x^+_1x^+_2-x^-_1x^-_2}
\eq\frac{g^2}{2x^+_1x^-_1x^+_2x^-_2}\,
\frac{x^+_1-x^+_2}{1-g^2/2x^-_1x^-_2}\,,\nln
B_{12}/S^0_{12}\eq -1
             +\frac{g^2}{2x^+_1x^-_1x^+_2x^-_2}\,
              \frac{x_1^+x_2^+-2x_1^-x_2^++x_1^-x_2^-}{1-g^2/2x^-_1x^-_2}\,
              \frac{x^+_1-x^+_2}{x_2^--x_1^+}\,,\\\nonumber
E_{12}/S^0_{12}\eq \frac{x_2^+-x_1^-}{x_2^--x_1^+}
         -\frac{g^2}{2x^+_1x^-_1x^+_2x^-_2}\,
          \frac{x_1^+x_2^+-2x_1^+x_2^-+x_1^-x_2^-}{1-g^2/2x^-_1x^-_2}\,
          \frac{x^+_1-x^+_2}{x_2^--x_1^+}\,,
\>
They can be derived
from the quadratic constraint \eqref{eq:Alg.xpmrel} 
between $x^+$ and $x^-$.

Let us compare to the results in \cite{Beisert:2005fw}
for the S-matrix in the $\alg{su}(1|2)$ sector of $\superN=4$ SYM. 
The S-matrix has manifest $\alg{su}(1|1)$ symmetry 
as explained in \cite{Beisert:2005wm}
and we obtain it by restricting to the
spin components $a,b,\alpha,\beta=1$.
Then only the elements 
$A,D,G,H,K,L$ in \tabref{tab:SCoeff} are relevant 
and the S-matrix agrees with \cite{Beisert:2005fw,Beisert:2005wm}.

\subsection{Properties}

We have already made use of the invariance of the S-matrix
in its construction
\[
\comm{\gen{J}_{1}+\gen{J}_{2}}{\smat\lvl{I}_{12}}=0.
\]
It however obeys a host of other important identities. 
First of all, it is an involution
\[
\smat\lvl{I}_{12}\smat\lvl{I}_{21}=1
\]
assuming that the undetermined phase obeys $S^0(x_1,x_2)S^0(x_2,x_1)=1$.
We have also verified that it satisfies the Yang-Baxter equation%
\footnote{M.~Staudacher has confirmed that the YBE is satisfied
at the first few perturbative orders in $g$.} 
\[\label{eq:YBE}
\smat\lvl{I}_{12}\smat\lvl{I}_{13}\smat\lvl{I}_{23}=\smat\lvl{I}_{23}\smat\lvl{I}_{13}\smat\lvl{I}_{12}.
\]
This is very tedious and we have made use of 
\texttt{Mathematica} to evaluate \eqref{eq:YBE}
on all three-particle states.
Note that the appearance of $\fldZ^\pm$ 
in \tabref{tab:SCoeff} can lead to additional phases
due to \eqref{eq:Alg.Shift}, e.g.
\[
\smat\lvl{I}_{12}\state{\phi_1\phi_2\psi_3}\lvl{I}
\to
\state{\psi_2\psi_1\fldZ^-\psi_3}\lvl{I}+\ldots
=
\frac{x_3^+}{x_3^-}\,\state{\psi_2\psi_1\psi_3\fldZ^-}\lvl{I}
+\ldots\,.
\]

It is also worth considering the $g=0$ limit
corresponding to one-loop gauge theory.
Here all the particle representations 
have central charge $C=\half$ and
transform as fundamentals under $\alg{su}(2|2)$.
When we set $\alpha=\order{g}$, it is easy to see that 
\[
\bigeval{\smat\lvl{I}_{12}}_{g=0}=\mathcal{P}^u_{12} S_{12}^0 
\lrbrk{\frac{u_2-u_1}{u_2-u_1-i}+\frac{i}{u_2-u_1-i}\,\mathcal{P}_{12}}
\]
where $\mathcal{P}_{12}$ is a graded permutation of the spin labels
$a,b,\alpha,\beta$
and $\mathcal{P}^u_{12}$ interchanges the spectral parameters $u_1,u_2$.
This agrees with the well-known S-matrix in the fundamental representation 
of $\alg{su}(2|2)$. We recover the
model found in \cite{Essler:1992aa}.

\subsection{Eigenstates}

A generic eigenstate $\state{\Psi}$ of 
the spin chain can now be represented by a
set of numbers $\set{x_1,\ldots,x_K}$ and a 
residual wave function $\state{\Psi\lvl{I}}$.
This residual wave function is given as a state of a new inhomogeneous spin chain 
with only four spin states $\set{\phi^1,\phi^2\mathpunct{|}\psi^1,\psi^2}$
such that spin site $k$ has momentum $p_k=p(x_k)$
along the original spin chain.
The eigenstate is 
\[
\state{\Psi}=\smat\lvl{I} \state{\Psi\lvl{I}}.
\]
Here $\smat\lvl{I}$ is the multi-particle S-matrix at level I.
In the case of infinitely many conserved charges, 
the set of momenta is preserved in the scattering process, 
i.e.~only the momenta can be permuted. 
Indications that this might be true were found
in \cite{Beisert:2003ys,Agarwal:2005jj}.
The S-matrix can thus be written as
\[
\smat\lvl{I}=\sum_{\pi\in S_K}\smat\lvl{I}_\pi.
\]
The S-matrix $\smat_\pi$ interchanges the sites and momenta of the spin chain 
$\state{\Psi\lvl{I}}$ according to the permutation $\pi$. 
If the S-matrix factorises, it can be written as a product 
over pairwise permutations of adjacent excitations 
\[\smat\lvl{I}_\pi=\prod_{(k,l)\in\pi}\smat\lvl{I}_{kl}.\]
Due to the YBE \eqref{eq:YBE} this product can be defined self-consistently. 
Let us therefore \emph{assume} that the S-matrix factorises
and that the Hamiltonian $\alg{C}$ is integrable.
An indirect verification of this assumption
is that the resulting Bethe equations
indeed reproduce several energies correctly \cite{Beisert:2005fw}.
This solves the problem of finding asymptotic eigenstates $\state{\Psi}$
of the infinite spin chain.

\section{Diagonalising the S-Matrix}
\label{sec:Diag}

The above solution for the infinite chain is complete,
but it requires a residual wave function $\state{\Psi\lvl{I}}$ to be specified. 
In other words, we have replaced the level-0 wave function $\state{\Psi}$ 
by a set of parameters $\set{x_1,\ldots,x_K}$ and a level-I wave function $\state{\Psi\lvl{I}}$.
We can now try to repeat this process and represent the spin chain 
$\state{\Psi\lvl{I}}$ by a set of parameters $\set{y_1,\ldots,y_{K'}}$
and a level-II wave function $\state{\Psi\lvl{II}}$.
This is the so-called nested Bethe ansatz
\cite{Yang:1967bm}.

\subsection{Vacuum}
\label{sec:Diag.Vac}

We start by choosing a level-II vacuum state consisting only of $\phi^1$'s
\[
\state{0}\lvl{II}=\state{\phi^1_1\ldots\phi^1_{K}}\lvl{I}.
\]
For any permutation $\pi$, the S-matrix $\smat\lvl{I}_\pi$ 
yields a total phase $S\lvl{I}_\pi$ 
times a vacuum of the inhomogeneous chain with permuted momenta
\[
\smat\lvl{I}_\pi\state{0}\lvl{II}=S\lvl{I}_\pi\state{0}\lvl{II}_\pi,\qquad
\state{0}\lvl{II}_\pi=\state{\phi^1_{\pi(1)}\ldots\phi^1_{\pi(K)}}\lvl{I}.
\]
The total phase is given by a product over two-particle phases
\[\label{eq:Diag.SPhase}
S\lvl{I}_\pi=\prod_{(k,l)\in\pi}S\lvl{I,I}(x_k,x_l),
\qquad
S\lvl{I,I}(x_k,x_l)=A(x_k,x_l)=S_0(x_k,x_l)\,\frac{x_k^--x_l^+}{x_k^+-x_l^-}\,.
\]
%

\subsection{Propagation}
\label{sec:Diag.Prop}

Now let us insert one excitation which might be of type
$\psi^1,\psi^2$ or $\phi^2$. 
If it is of type $\psi^1$ or $\psi^2$, 
an action of the S-matrix shifts this excitation around.
If it is of type $\phi^2$, however, the S-matrix
can shift it around, but it can also convert it into one
excitation of type $\psi^1$ and $\psi^2$ each.
Subsequently, these two will be propagated by the S-matrix on an 
individual basis. 
Therefore is $\phi^2$ a non-elementary double excitation
whereas $\psi^1,\psi^2$ are the only two elementary excitations
of the vacuum $\state{0}\lvl{II}$.

A generic one-excitation state is given by 
\[\label{eq:Diag.OneEx}
\state{\psi^\alpha}\lvl{II}=\sum_{k=1}^K 
\Psi_k(y)\,\state{\phi^1_1\ldots\psi^{\alpha}_k\ldots\phi^1_K}\lvl{I}
\]
with some wave function $\Psi_k(y)$.
For this wave function we make a plane wave ansatz in 
the inhomogeneous background which is determined through the $x_l$'s
\[
\Psi_k(y)=f(y,x_k)\prod_{l=1}^{k-1} S\lvl{II,I}(y,x_{l}).
\]
Here $S\lvl{II,I}(y,x_{k'})$ represents the phase 
when permuting the excitation past a background field
and $f(y,x_k)$ is a factor for the combination of the
excitation with the background field.

We demand compatibility of the wave function with the S-matrix.
This means that $\smat\lvl{I}_\pi$ merely multiplies
the state by the above $S\lvl{I}_\pi$ in 
\eqref{eq:Diag.SPhase}
and permutes the momenta
\[
\smat\lvl{I}_\pi\state{\psi^\alpha}\lvl{II}=
S\lvl{I}_\pi \state{\psi^\alpha}\lvl{II}_\pi,
\qquad
\state{\psi^\alpha}\lvl{II}_{\pi}=
\sum_{k=1}^K
\Psi_{\pi,k}(y)\,\state{\phi^1_{\pi(1)}\ldots\psi^{\alpha}_{\pi(k)}\ldots\phi^1_{\pi(K)}}\lvl{I}.
\]
with
\[
\Psi_{\pi,k}(y)=f(y,x_{\pi(k)})\prod_{l=1}^{k-1} S\lvl{II,I}(y,x_{\pi(l)}).
\]
To solve this problem, 
it is sufficient to consider a spin chain with only two sites
\<
\state{\psi^\alpha}\lvl{II}\eq
f(y,x_1)\state{\psi^{\alpha}_1\phi^1_2}\lvl{I}
+f(y,x_2)S\lvl{II,I}(y,x_1)\,\state{\phi^1_1\psi^{\alpha}_2}\lvl{I},
\nln
\state{\psi^\alpha}\lvl{II}_{\pi}\eq
f(y,x_2)\state{\psi^{\alpha}_2\phi^1_1}\lvl{I}
+f(y,x_1)S\lvl{II,I}(y,x_2)\,\state{\phi^1_2\psi^{\alpha}_1}\lvl{I}.
\>
We thus demand
\[
\smat\lvl{I}_{12}\state{\psi^\alpha}\lvl{II}=S\lvl{I,I}(x_1,x_2)\,\state{\psi^\alpha}\lvl{II}_{\pi}
\]
which amounts to 
\<
f(y,x_1)\,K(x_1,x_2)+f(y,x_2)\,S\lvl{II,I}(y,x_1)\,G(x_1,x_2) \eq f(y,x_2)\, A(x_1,x_2),
\\\nn
f(y,x_1)\,L(x_1,x_2)+f(y,x_2)\,S\lvl{II,I}(y,x_1)\, H(x_1,x_2) \eq f(y,x_1)\,S\lvl{II,I}(y,x_2)\,A(x_1,x_2).
\>
These two equations are solved by
\[
S\lvl{II,I}(y,x_k)=\frac{y-x^-_k}{y-x^+_k}\,,\qquad
f(y,x_k)=\frac{y\gamma_k}{y-x^+_k}\,.
\]
%

\subsection{Scattering}
\label{sec:Diag.Scat}

For a two-excitation state we make an ansatz of 
two superimposed plane waves
\[
\state{\psi_1^\alpha\psi_2^\beta}\lvl{II}=\sum_{k<l=1}^K 
\Psi_k(y_1)\Psi_l(y_2)
\,\state{\phi^1_1\ldots\psi^{\alpha}_k\ldots\psi^{\beta}_l\ldots\phi^1_K}\lvl{I}.
\]
This solves the compatibility condition
$\smat\lvl{I}_\pi\state{\psi_1^\alpha\psi_2^\beta}\lvl{II}
=S\lvl{I}_\pi \state{\psi_1^\alpha\psi_2^\beta}\lvl{II}_\pi$
except when the two excitation are neighbours.
We should also consider a state with
one excitation $\phi^2\fldZ^+$ which can undergo mixing with the
previous state
\[\label{eq:Diag.Coincident}
\state{\phi^2_{12}\fldZ^+}\lvl{II}=\sum_{k=1}^K 
\Psi_k(y_1)\Psi_k(y_2)f(y_1,y_2,x_k)
\,\state{\phi^1_1\ldots\phi^{2}_k\fldZ^+\ldots\phi^1_K}\lvl{I}.
\]
Here, $f(y_1,y_2,x_k)$ represents a factor which occurs
when two excitations reside on the same site.
A generic two excitation eigenstate must be of the form
\[
\state{\Psi\lvl{II}}=
\state{\psi_1^\alpha\psi_2^\beta}\lvl{II}
+
\varepsilon^{\alpha\beta}\state{\phi^2_{12}\fldZ^+}\lvl{II}
+
\smat\lvl{II}_{12}
\state{\psi_1^\alpha\psi_2^\beta}\lvl{II}
\]
with an $\alg{su}(2)$ symmetric S-matrix 
\[
\smat\lvl{II}_{12}
\state{\psi_1^\alpha\psi_2^\beta}\lvl{II}=
M_{12}\state{\psi_2^\alpha\psi_1^\beta}\lvl{II}
+N_{12}\state{\psi_2^\beta\psi_1^\alpha}\lvl{II}.
\]
Again we impose the compatibility condition
\[\label{eq:Diag.Compt2}
\smat\lvl{I}_{\pi}
\state{\Psi\lvl{II}}
=
S_{\pi}
\state{\Psi\lvl{II}}_\pi
\]
which is trivially satisfied when the two excitations are not
neighbours. To solve the relation exactly we need to consider 
only a two-site state
\<
\state{\Psi\lvl{II}}\eq
f(y_1,x_1)f(y_2,x_2)S\lvl{II,I}(y_2,x_1)
\,\state{\psi^{\alpha}_1\psi^{\beta}_2}\lvl{I}
\nl
+
f(y_1,x_1)f(y_2,x_1)f(y_1,y_2,x_1)(x_2^-/x_2^+)
\,\varepsilon^{\alpha\beta}\state{\phi^{2}_1\phi^1_2\fldZ^+}\lvl{I}
\nl
+
f(y_1,x_2)f(y_2,x_2)S\lvl{II,I}(y_1,x_1)S\lvl{II,I}(y_2,x_1)f(y_1,y_2,x_2)
\,\varepsilon^{\alpha\beta}\state{\phi^1_1\phi^{2}_2\fldZ^+}\lvl{I}
\nl
+
M(y_1,y_2)
f(y_2,x_1)f(y_1,x_2)S\lvl{II,I}(y_1,x_1)
\,\state{\psi^{\alpha}_1\psi^{\beta}_2}\lvl{I}
\nl
+N(y_1,y_2)
f(y_2,x_1)f(y_1,x_2)S\lvl{II,I}(y_1,x_1)
\,\state{\psi^{\beta}_1\psi^{\alpha}_2}\lvl{I}
\>
and the state $\state{\Psi\lvl{II}}_\pi$ 
where $x_1$ and $x_2$ are interchanged.
We find the unique solution of \eqref{eq:Diag.Compt2}
\<
M_{12}\eq -\frac{i}{y_1+g^2/2y_1-y_2-g^2/2y_2+i}=-\frac{i}{v_1-v_2+i}\,,
\nln
N_{12}\eq-\frac{y_1+g^2/2y_1-y_2-g^2/2y_2}{y_1+g^2/2y_1-y_2-g^2/2y_2+i}=
-\frac{v_1-v_2}{v_1-v_2+i}\,,
\>
where the new spectral parameter $v_k$ is related to $y_k$ as
\[
v_k=y_k+\frac{g^2}{2y_k}\,.
\]
The factor for two coincident excitations in \eqref{eq:Diag.Coincident} is
\[\label{eq:Diag.CoincidentFactor}
f(y_1,y_2,x_k)=
\frac{\alpha}{\gamma_k^2}\,
\frac{x^-_k-x^+_k}{x^+_k}\,
\frac{y_1y_2-x^-_kx^+_k}{y_1y_2}\,
\frac{y_1-y_2}{y_1+g^2/2y_1-y_2-g^2/2y_2+i}\,.
\]
In \appref{sec:Alt2} we will 
present an alternative notation for wave functions
which is somewhat more transparent
and should naturally generalise to more than
two excitations.

\subsection{Final Level}

The level-II S-matrix $\smat\lvl{II}_{12}$ has the standard
form of a $\alg{su}(2)$ invariant S-matrix with spectral
parameters $v_k=y_k+g^2/2y_k$. 
It is therefore clear that the remaining elements 
of the diagonalised S-matrix are%
\footnote{Note that the excitation of type II is fermionic.
For the diagonalised S-matrix we shall use the convention that
scattering of two fermions introduces an additional factor of $-1$.
Hence $S\lvl{II,II}=1$.}
\<
S\lvl{II,II}(y_1,y_2)\eq -M(y_1,y_2)-N(y_1,y_2)=1,
\nln
S\lvl{III,II}(w_1,y_2)\eq 
\frac{w_1-y_2-g^2/2y_2-\ihalf}{w_1-y_2-g^2/2y_2+\ihalf}
=
\frac{w_1-v_2-\ihalf}{w_1-v_2+\ihalf}\,,
\nln
S\lvl{III,III}(w_1,w_2)\eq \frac{w_1-w_2+i}{w_1-w_2-i}\,.
\>
Eigenstates of the Hamiltonian are now determined
through a set of main parameters $\set{x_1,\ldots,x_{K\lvl{I}}}$
as well as several auxiliary parameters $\set{y_1,\ldots,y_{K\lvl{II}}}$
and $\set{w_1,\ldots,w_{K\lvl{III}}}$.
The spin chain picture has completely dissolved.

\subsection{Bethe Equations}

Bethe equations are periodicity conditions 
for a state of the original spin chain.
As the length fluctuates, we cannot define the period,
but if we also impose cyclicity this is not a problem.
The generic Bethe equations for a diagonalised
S-matrix $S^{AB}(x^A_k,x^B_l)$ read
\[
1=\mathop{\prod_{B=0}\lvl{III}\prod_{l=1}^{K^B}}_{(B,l)\neq(A,k)} S^{BA}(x^B_l,x^A_k).
\]
Here $K^A$ is the number of excitations of type $A\in\set{0,\mathrm{I},\mathrm{II},\mathrm{III}}$. 
So far we have not introduced the quasi-excitations of type 0: 
These are sites of the original spin chain 
and they do not carry an individual momentum parameter
for this homogeneous spin chain. 
They only scatter with excitations of type I defining the wave function of a
homogeneous plane wave
\[
S\lvl{I,0}(x_k,\cdot)=\frac{x^+_k}{x^-_k}=e^{ip_k}.
\]
Imposing a Bethe equation at level 0 implies
that sites can be permuted around the chain without a net phase shift.
This operation is a global shift
and invariance is equivalent to the zero-momentum condition \eqref{eq:Alg.Phys}, 
i.e.~the physicality constraint 
$\gen{P}\state{\Psi}=\gen{K}\state{\Psi}=0$ in \secref{sec:Alg.Dyn}.
Clearly, the S-matrix satisfies the involution condition
\[
S^{A,B}(x^A_k,x^B_l)=1/S^{B,A}(x^B_l,x^A_k)
\]
from which the remaining matrix elements can be read off.
The asymptotic Bethe equations are summarised in \tabref{tab:Bethe}.
Here $v_k=y_k+g^2/2y_k$ and $x^\pm_k$ are related by
\eqref{eq:Alg.xpmrel}.%

\begin{table}
\<
1 \eq \prod_{l=1}^{K\lvl{I}}\frac{x^+_l}{x^-_l}\,,
\nln
1 \eq \lrbrk{\frac{x^-_k}{x^+_k}}^{K\lvl{0}}
\prod_{\textstyle\atopfrac{l=1}{l\neq k}}^{K\lvl{I}} 
\lrbrk{S^0(x_l,x_k)\,\frac{x^+_k-x^-_l}{x^-_k-x^+_l}}
\prod_{l=1}^{K\lvl{II}} \frac{x^-_k-y_l}{x^+_k-y_l}\,,
\nln
1 \eq 
\prod_{l=1}^{K\lvl{I}} \frac{y_k-x^+_l}{y_k-x^-_l}
\prod_{l=1}^{K\lvl{III}} \frac{v_k-w_l+\ihalf}{v_k-w_l-\ihalf}\,,
\nln
1 \eq 
\prod_{l=1}^{K\lvl{II}} \frac{w_k-v_l+\ihalf}{w_k-v_l-\ihalf}
\prod_{\textstyle\atopfrac{l=1}{l\neq k}}^{K\lvl{III}} 
\frac{w_k-w_l-i}{w_k-w_l+i}\,.\nn
\>
\caption{Asymptotic Bethe equations for
the dynamic $\alg{su}(2|3)$ spin chain.}
\label{tab:Bethe}
\end{table}

In order to understand the number of excitations $K^A$,
we first of all convert all fields $\phi^2$ into 
$\psi^1\psi^2/\fldZ\phi^1$. Then we follow through
the above nested Bethe ansatz and find
\[\begin{array}{rclclcl}
  K\lvl{0}\eq N(\fldZ)+N(\phi^1)+N(\psi^1)+N(\psi^2)
          \eq p+2q+2r-s
          \eq r_1+2r_2-r_3,
\\[3pt]
  K\lvl{I}\eq N(\phi^1)+N(\psi^1)+N(\psi^2)+N(\phi^2)
          \eq q+2r-s
          \eq r_2+r_4,
\\[3pt]
 K\lvl{II}\eq N(\psi^1)+N(\psi^2)+2N(\phi^2)
          \eq 2r-s
          \eq r_3+2r_4,
\\[3pt]
K\lvl{III}\eq N(\psi^2)+N(\phi^2)
          \eq r-s
          \eq r_4.
\end{array}
\]
Here, $[p,q;r+\half \delta D;s]$ are the Dynkin labels 
of the state when the Dynkin diagram is \mbox{O--O--X--O} and 
$[r_1;r_2+\half \delta D;r_3;r_4+\half \delta D]$ 
are the Dynkin labels when the diagram is \mbox{O--X--O--X}. 
These are related by $p=r_1,q=r_2-r_3-r_4;s=r_3;r=r_3+r_4$.
Note that the highest-weight state in a multiplet is determined 
using the Dynkin diagram \mbox{O--X--O--X}.

The derived Bethe equations agree with the equations
conjectured in \cite{Beisert:2005fw}. To
see this, we first eliminate the flavours $1,2,3$ to
restrict to the $\alg{su}(2|3)$ sector. Then we trade in 
all Bethe roots of type $7$ for Bethe roots of type
$5$ by means of the duality transformation.
Finally, we identify flavours I,II,III with $4,5,6$,
respectively. In other words, the Bethe roots
$x,u,y,v,w$ correspond to $x_4,u_4,x_5,u_5,u_6$.

\subsection{Symmetry Enhancement}

Superficially, the Bethe equations in \tabref{tab:Bethe} 
look as though they originate from 
the Dynkin diagram \mbox{O--X--O}, i.e.~a spin chain 
with $\alg{su}(2|2)$ symmetry.
However, the full symmetry algebra of the considered
spin chain is $\alg{su}(3|2)$ by construction.
This means that some of the symmetry must be hidden.

Symmetries in the Bethe equations are represented
by Bethe roots at special positions,
conventionally at $\infty$.
Indeed, one can add a Bethe root 
$x^\pm\to \infty$ (flavour I), 
$y,v\to \infty$ (flavour II) or 
$w\to \infty$ (flavour III) 
to any existing set of Bethe roots.
If the original set satisfies the Bethe equations, 
the new set does so as well, because the scattering
between these special excitations and any other
excitation is trivial, $S=1$.

Symmetry enhancement for the Bethe equations in \tabref{tab:Bethe}
works as follows: One adds a Bethe root $y=0$ and removes
a quasi-excitation of type $0$ at the same time.
In the Bethe equation for $x_k$, the
effect of adding $y=0$ and removing a 
quasi-excitation cancels. In the Bethe
equation for a $w_k$, the scattering with
$y=0$ is trivial, because $v=y+g^2/2y=\infty$.
The equation for some other $y_k$ is not modified
due to the absence of self-scattering terms for fermions.
Finally, in the equation for $y$ itself,
the net scattering with all $x_l$'s is equivalent 
to the zero-momentum condition \eqref{eq:Alg.Phys}. The latter is
effectively the Bethe equation for the
(removed) quasi-excitation.

In conclusion, the Bethe equations have a hidden $\alg{su}(2|3)$ symmetry. 
This however requires that the physicality constraint holds.
One can also derive the S-matrix and Bethe equations assuming that 
the residual symmetry at level I is $\alg{su}(1|2)$. 
This avenue is considered in \appref{sec:SMatrix.Alt}.

\subsection{Abelian Phase}
\label{sec:Diag.Phase}

We have solved the asymptotic spectrum of the $\alg{su}(2|3)$ dynamic 
spin chain \cite{Beisert:2003ys} up to the overall function
$S_0(x_k,x_l)$.
The analysis of a similar class of long-range spin chains in
\cite{Beisert:2005wv} has produced a suggestive generic form for this
function. 
Clearly, it does not necessarily have to apply to this particular spin chain, 
but it is worth contemplating the possibility. 
Here is summary of the results of \cite{Beisert:2005wv}:
The overall factor is
\[
S_0(x_l,x_k)=
\frac{1-g^2/2x^+_{k}x^-_{l}}{1-g^2/2x^-_{k}x^+_{l}}\,
\exp \bigbrk{2i\theta_{kl}} 
\]
with the dressing phase
\[
\theta_{kl}=
\sum_{r=2}^\infty
\sum_{s=r+1}^\infty
\beta_{rs}(g^2)
\bigbrk{q_{r,k}\,q_{s,l}-q_{s,k}\,q_{r,l}}.
\]
and the $r$-th moment of the $k$-th excitation
\[
q_{r,k}=\frac{1}{r-1}\lrbrk{\frac{i}{(x^+_k)^{r-1}}-\frac{i}{(x^-_k)^{r-1}}}  .
\]
The coefficient functions $\beta_{rs}(g)$, $r>s$, can be chosen freely, 
but the structure of the algebra generators imposes some constraints:
Compatibility with the range of the interactions requires
$\beta_{rs}(g^2)=\order{g^{2s-2}}$.
Compatibility with gauge theory Feynman diagrams imposes
the more restrictive constraint
$\beta_{rs}(g^2)=\order{g^{2r+2s-4}}$.
Finally, the coefficients $\beta_{rs}$ with odd $r+s$ violate parity.
The author believes that all these coefficients can be 
realised by the underlying $\alg{su}(2|3)$ spin chain. 
In the analysis of \cite{Beisert:2003ys} only
the first, $\beta_{23}(g^2)$ can be seen at $\order{g^4}$. 

In \cite{Beisert:2005wv} two further sequences 
of parameters related to propagation 
and mixing of charges were identified.
Here, these degrees of freedom are fixed by the structure of the algebra,
cf.~\eqref{eq:Alg.xumap},
and the inclusion of the Hamiltonian in the algebra. 
Finally, we note that \eqref{eq:Alg.xumap} is not
the correct map for the Inozemtsev spin chain \cite{Inozemtsev:1989yq,Inozemtsev:2002vb}, 
cf.~the appendix of \cite{Beisert:2004hm}.
This proves that the Inozemtsev spin chain cannot 
be an accurate description of the $\alg{su}(2)$ sector
of planar $\superN=4$ SYM beyond three loops
which remained as a logical possibility after \cite{Serban:2004jf}.
Conversely, we cannot guarantee that the spin chain of \cite{Beisert:2004hm} 
is the correct (asymptotic) description at starting from four loops; 
proper scaling in the thermodynamic limit may be violated in 
other ways or even integrability might break 
(although the latter does not seem likely).

\section{Generalisation to $\alg{psu}(2,2|4)$ and $\superN=4$ SYM}
\label{sec:N4}

In $\superN=4$ super Yang-Mills, there are $(8|8)$ types
of level-I excitations \cite{Berenstein:2002jq}. 
These transform under the residual algebra $\alg{psu}(2|2)^2\ltimes \Reals^3$
\cite{Beisert:2004ry}.%
\footnote{A very similar algebra appeared in 
the study of mass deformed M2 branes 
\cite{Pope:2003jp,Bena:2004jw,Lin:2005nh}. 
It would be interesting to find out if there is a deeper connection. 
Also the residual algebra $\alg{su}(2|2)\ltimes \Reals^2$ 
for the $\alg{su}(2|3)$ sector of $\superN=4$ SYM appears
to play a for M5 branes \cite{Lin:2005nh}.}
The generators of the bosonic subalgebra 
$\alg{su}(2)^4$ are $\gen{L},\gen{R},\gen{\dot L},\gen{\dot R}$,
the fermionic generators are $\gen{Q},\gen{S},\gen{\dot Q},\gen{\dot S}$.
The dotted algebra relations are the same as for the
undotted ones \eqref{eq:Alg.CommSU22,eq:Alg.CommExt} with the
central charges shared among the two algebras
$(\gen{\dot C},\gen{\dot K},\gen{\dot P})=(\gen{C},\gen{K},\gen{P})$.
The set of $(8|8)=(2|2)\times (2|2)$ excitations now transforms under each 
extended $\alg{psu}(2|2)\ltimes \Reals^3$ subalgebra as $(2|2)$
in \secref{sec:Alg.Rep}.
The $(8|8)$ composite fields are of four types: 
$(\phi\dot\phi)$ is a quartet of scalars, 
$(\phi\dot\psi)$ and $(\psi\dot\phi)$ are two
quartets of fermions and 
$(\psi\dot\psi)$ is a quartet of covariant derivatives.

We can now apply the above results 
for the algebra, S-matrix and Bethe equations 
to $\superN=4$ SYM.
Due to invariance under each factor of the residual symmetry, 
the S-matrix should be
\[
\smat^{\superN=4}_{kl}=\smat\lvl{I}_{kl}\,\dot\smat\lvl{I}_{kl}/A_{kl}.
\]
with some overall undetermined phase $S_0(x_k,x_l)$,
c.f.~the remarks in \secref{sec:Diag.Phase}.
Similarly, the asymptotic Bethe equations can be composed from 
those in \tabref{tab:Bethe}.
Here, the main Bethe roots $x^\pm_k$ are
shared among the two sectors, but the 
auxiliary Bethe roots $y_k,w_k$
are duplicated $\dot y_k,\dot w_k$.
The complete Bethe equations are as
in \cite{Beisert:2005fw}.

\section*{Acknowledgements}

I would like to thank Matthias Staudacher for 
his initial collaboration on this project.
I would also like to thank 
Juan Maldacena, 
Nicolai Reshetikhin,
Radu Roiban
and Yao-Zhong Zhang
for useful discussions. 
This work is supported in part by
the U.S.~National Science Foundation Grant No.~PHY02-43680. 
I thank KITP for hospitality during part of the work
and acknowledge partial support of NSF grant PHY99-07949 while at KITP. 
Any opinions, findings and conclusions or recommendations expressed in
this material are those of the author and do not necessarily
reflect the views of the National Science Foundation.

\appendix

\section{A Contraction of $\alg{d}(2,1;\epsilon)$}
\label{sec:contract}

The exceptional superalgebra $\alg{d}(2,1;\epsilon)$ consists
of three triplets of $\alg{su}(2)$ generators
$\gen{J}^a{}_b$, $\gen{J}^\alpha{}_\beta$,
$\gen{J}^{\mathfrak{a}}{}_{\mathfrak{b}}$
and an octet of fermionic generators $\gen{J}^{a\beta\mathfrak{c}}$.
The $\alg{su}(2)^3$ generators commute canonically
\<
\comm{\gen{J}^a{}_b}{\gen{J}^c{}_d}
\eq \delta_b^c\gen{J}^a{}_d
   -\delta^a_d\gen{J}^c{}_b,
\nln
\comm{\gen{J}^\alpha{}_\beta}{\gen{J}^\gamma{}_\delta}
\eq \delta_\beta^\gamma\gen{J}^\alpha{}_\delta
   -\delta^\alpha_\delta\gen{J}^\gamma{}_\beta,
\nln
\comm{\gen{J}^{\mathfrak{a}}{}_{\mathfrak{b}}}{\gen{J}^{\mathfrak{c}}{}_{\mathfrak{d}}}
\eq \delta_{\mathfrak{b}}^{\mathfrak{c}}\gen{J}^{\mathfrak{a}}{}_{\mathfrak{d}}
   -\delta^{\mathfrak{a}}_{\mathfrak{d}}\gen{J}^{\mathfrak{c}}{}_{\mathfrak{b}}.
\>
The fermionic generators transform
in the fundamental representation of each 
$\alg{su}(2)$ factor
\<
\comm{\gen{J}^a{}_b}{\gen{J}^{c\delta\mathfrak{e}}}
\eq \delta_b^c\gen{J}^{a\delta\mathfrak{e}}
   -\half\delta_b^a\gen{J}^{c\delta\mathfrak{e}},
\nln
\comm{\gen{J}^\alpha{}_\beta}{\gen{J}^{c\delta\mathfrak{e}}}
\eq \delta_\beta^\delta\gen{J}^{c\alpha\mathfrak{e}}
   -\half\delta_\beta^\alpha\gen{J}^{c\delta\mathfrak{e}},
\nln
\comm{\gen{J}^{\mathfrak{a}}{}_{\mathfrak{b}}}{\gen{J}^{c\delta\mathfrak{e}}}
\eq \delta_{\mathfrak{b}}^{\mathfrak{e}}\gen{J}^{c\delta\mathfrak{a}}
   -\half\delta_{\mathfrak{b}}^{\mathfrak{a}}\gen{J}^{c\delta\mathfrak{e}}.
\>
Finally, the anticommutator of the fermionic generators is 
\[
\acomm{\gen{J}^{a\beta\mathfrak{c}}}{\gen{J}^{d\epsilon\mathfrak{f}}}
=
\alpha\,\varepsilon^{ak}\varepsilon^{\beta\epsilon}\varepsilon^{\mathfrak{c}\mathfrak{f}}\gen{J}^d{}_k
+\beta\,\varepsilon^{ad}\varepsilon^{\beta\kappa}\varepsilon^{\mathfrak{c}\mathfrak{f}}\gen{J}^\epsilon{}_\kappa
+\gamma\,\varepsilon^{ad}\varepsilon^{\beta\epsilon}\varepsilon^{\mathfrak{c}\mathfrak{k}}\gen{J}^{\mathfrak{f}}{}_{\mathfrak{k}}.
\]
The Jacobi identity requires $\alpha+\beta+\gamma=0$
and a rescaling of $\gen{J}^{a\beta\mathfrak{c}}$ leads
to a rescaling of $(\alpha,\beta,\gamma)$.
The parameter of $\alg{d}(2,1;\epsilon)$ is
given by $\epsilon=\gamma/\alpha$ or any other
of the six quotients made from two 
of the coefficients $\alpha,\beta,\gamma$.

We now derive the algebra in \secref{sec:Alg.Alg} 
as a contraction of the above algebra.
First of all we identify two
of the $\alg{su}(2)$'s
\[
\gen{J}^a{}_b=\gen{R}^a{}_b,\qquad
\gen{J}^\alpha{}_\beta=\gen{L}^\alpha{}_\beta.
\]
The third $\alg{su}(2)$ will be contracted, we split up 
the generator $\gen{J}^{\mathfrak{a}}{}_{\mathfrak{b}}$ as follows
\[
\gen{J}^1{}_2=\epsilon^{-1}\,\gen{P},\qquad
\gen{J}^1{}_1=-\gen{J}^2{}_2=-\epsilon^{-1}\,\gen{C},\qquad
\gen{J}^2{}_1=-\epsilon^{-1}\,\gen{K}.
\]
The fermionic generator yields the supersymmetry generators
\[
\gen{J}^{a\beta 1}=\varepsilon^{ac}\gen{Q}^\beta{}_{c},\qquad
\gen{J}^{a\beta 2}=\varepsilon^{\beta\gamma}\gen{S}^a{}_{\gamma}.
\]
Finally, the three constants of the exceptional algebra are adjusted
to $\alg{d}(2,1;\epsilon)$
\[
\alpha=-1-\epsilon,\qquad
\beta=1,\qquad
\gamma=\epsilon.
\]
Sending $\epsilon\to 0$ leads to the commutation relations
in \secref{sec:Alg.Alg}.

\section{Alternative Notation with $\alg{su}(1|2)$ Symmetry}
\label{sec:SMatrix.Alt}

The manifest symmetry of the Bethe equations 
is $\alg{su}(1|2)$, i.e.~the 
residual symmetry at level I appears to be 
$\alg{su}(1|2)$ and not $\alg{su}(2|2)$.
In fact, we can work with $\alg{su}(1|2)$
as the manifest symmetry of the S-matrix and thereby avoid 
the effects of a fluctuating length.
Let us outline this picture here.

We first define the two bosonic excitations 
as $\phi:=\phi^1$ and $\chi:=\phi^2\fldZ^+$.
Then the multiplet
$(\phi\mathpunct{|}\psi^1,\psi^2\mathpunct{|}\chi)$
transforms in a typical representation 
$\rep{(1|2|1)}$ of $\alg{su}(1|2)$.
This representation is like the one discussed in 
\secref{sec:Alg.Rep} but the index $a$ is restricted to the value $1$.
There is no complication from a fluctuating length as
in \eqref{eq:Rep.ActionQS} for $\gen{Q}^\alpha$ transforms 
$\phi=\phi^1$ to $\psi^\alpha$ and $\phi^\beta$ to
$\varepsilon^{\alpha\beta}\phi^2\fldZ^+=\varepsilon^{\alpha\beta}\chi$.
Similarly, $\gen{S}_\alpha$ transforms between 
$(\phi\mathpunct{|}\psi^1,\psi^2\mathpunct{|}\chi)$ 
in the opposite direction.
The spin chain becomes static.
Note that for an excitation with central charge $C=+\half$,
the representation splits in two parts
$\rep{(1|2|0)}$ and $\rep{(0|0|1)}$, 
i.e.~a fundamental and a trivial representation.
This is the common breaking pattern for typical
representations of $\alg{su}(2|1)$.

To understand the possible degrees of freedom of an invariant S-matrix 
one should investigate the irreducible representations
in the tensor product $\rep{(1|2|1)}^2$ \cite{Bracken:1994hz}.
There are three irreps which could be described by the symbols 
$\rep{(1|2|1|0|0)}$, $\rep{(0|2|4|2|0)}$ and $\rep{(0|0|1|2|1)}$.
The S-matrix thus acts on selected representatives as
\<\label{eq:Ssu12}
\smat\lvl{I}_{12}\,\state{\phi_1\phi_2}\lvl{I}\eq
S^1_{12}
\,\state{\phi_2\phi_1}\lvl{I},
\nln
\smat\lvl{I}_{12}\,\state{\psi^{\{\alpha}_1\psi^{\beta\}}_2}\lvl{I}\eq
-S^2_{12}\,\state{\psi^{\{\alpha}_2\psi^{\beta\}}_1}\lvl{I},
\nln
\smat\lvl{I}_{12}\,\state{\chi_1\chi_2}\lvl{I}\eq
S^3_{12}\,\state{\chi_2\chi_1}\lvl{I},
\>
the action on the other states is determined through $\alg{su}(1|2)$ invariance.
From symmetry arguments alone 
the three factors $S^k_{12}$ are independent. 
It is, however, very likely that they are interrelated
by the Yang-Baxter equation \eqref{eq:YBE}.

In the main text we have used 
invariance under $\alg{su}(2|2)\ltimes \Reals^2$
to relate the coefficients and found 
\<\label{eq:Ssu12coeff}
S^1_{12}\eq
S^0_{12}\,\frac{x^+_2-x^-_1}{x^-_2-x^+_1}\,,
\nln
S^2_{12}\eq S^0_{12},
\nln
S^3_{12}\eq
S^0_{12}\,\frac{g^2/2x^+_2-g^2/2x^-_1}{g^2/2x^-_2-g^2/2x^+_1}
=
S^0_{12}\,\frac{x^-_2}{x^+_2}\,\frac{x^+_1}{x^-_1}\,
\frac{x^+_2-x^-_1}{x^-_2-x^+_1}\,.
\>
It is straightforward to see that this S-matrix agrees
with \tabref{tab:SCoeff}. 
For the last line in \eqref{eq:Ssu12,eq:Ssu12coeff} 
one should note that $\chi=\phi^2\fldZ^+$ requires 
the introduction of factors of $\exp(ip)=x^+/x^-$ 
due to shifts of $\fldZ^+$ \eqref{eq:Alg.Shift}.

Spin chains with the same symmetry group and 
the same type of representation have been 
investigated in \cite{Bracken:1994hz,Bracken:1995aa}.
The above expressions \eqref{eq:Ssu12coeff} for 
the eigenvalues of the S-matrix however do not 
agree with the expressions in \cite{Bracken:1994hz,Bracken:1995aa}.
Also the Bethe equations for the same model in \cite{Pfannmuller:1996vp,Ramos:1996my}
are incompatible with our equations in \tabref{tab:Bethe}.
The results in \cite{Bracken:1994hz,Bracken:1995aa,Pfannmuller:1996vp,Ramos:1996my}
are certainly correct and it seems that
\eqref{eq:Ssu12coeff} is an exceptional solution of the YBE. 
The existence of such a solution
might be attributed to the fact that the 
representation of the excitations is correlated to the momentum
by \eqref{eq:Alg.Dispersion}, see also \cite{Beisert:2005wm}.
The distinction to the $\alg{su}(1|1)$ case in \cite{Beisert:2005wm}
appears to be that we cannot use an arbitrary dispersion relation, 
but only \eqref{eq:Alg.Dispersion} is valid.
It would be useful to understand the derivation 
with manifest $\alg{su}(2|1)$ symmetry better.

\section{Using Generators to Construct Level-II States}
\label{sec:Alt2}

In \secref{sec:Diag.Vac,sec:Diag.Prop,sec:Diag.Scat}
we have determined the diagonalised wave functions 
of two level-II excitations. 
Here we will present an alternative notation 
which easily generalises to more than two level-II excitations.
This ansatz makes use of the supersymmetry generators 
$(\gen{Q}^\alpha{}_1)_k$ to create an excitation 
$\psi^\alpha$ from the vacuum of $\phi^1$'s
\<
(\gen{Q}^\alpha{}_1)_k\,\state{0}\lvl{II}\eq
a_k\state{\phi^1_1\ldots\psi^{\alpha}_k\ldots\phi^1_K}\lvl{I},
\nln
(\gen{Q}^\alpha{}_1)_k(\gen{Q}^\beta{}_1)_l \state{0}\lvl{II}
\eq a_ka_l\state{\phi^1_1\ldots\psi^{\alpha}_k\ldots\psi^{\beta}_l\ldots\phi^1_K}\lvl{I},
\nln
(\gen{Q}^\alpha{}_1)_k(\gen{Q}^\beta{}_1)_k \state{0}\lvl{II}
\eq
a_kb_k \varepsilon^{\alpha\beta}\state{\phi^1_1\ldots\phi^{2}_k\fldZ^+\ldots\phi^1_K}\lvl{I}.
\>
The advantage of this notation is that 
various factors from the algebra, such as $a_k,b_k$, will be absorbed into the 
application of the symmetry generators.
The single-excitation state in 
\eqref{eq:Diag.OneEx} will now be written in a slightly 
different way
\[
\state{\psi^\alpha}\lvl{II}=\sum_{k=1}^K 
\frac{x^-_k \Psi_{k-1}(y)-x^+_k\Psi_{k}(y)}{x^-_k-x^+_k}\,(\gen{Q}^\alpha{}_1)_k\,\state{0}\lvl{II},
\qquad
\Psi_k(y)=\prod_{l=1}^{k} S\lvl{II,I}(y,x_{l}).
\]
Being somewhat sloppy about the terms at $k=0,K$
we can rewrite the one-excitation state as
\[\label{eq:Alt2.OneEx}
\state{\psi^\alpha}\lvl{II}=\sum_{k=0}^K 
\Psi_{k}(y)\bigbrk{(\gen{Q}^\alpha{}_1)^-_{k}+(\gen{Q}^\alpha{}_1)^+_{k+1}}\state{0}\lvl{II}.
\]
Here we have introduced the dressed generators
\[
(\gen{Q}^\alpha{}_1)^\pm_{k}=\frac{x^\mp_{k}}{x^\mp_{k}-x^\pm_{k}}\,(\gen{Q}^\alpha{}_1)_{k}.
\]
The formula \eqref{eq:Alt2.OneEx} can now be interpreted as follows:
The level-II excitation $y$ is permuted along the level-I chain
using the scattering phase $S\lvl{II,I}$ until it
is between $x_{k}$ and $x_{k+1}$. At this point it can
be joined with the vacuum either to the left by $(\gen{Q}^\alpha{}_1)^-_{k}$ 
or to the right by $(\gen{Q}^\alpha{}_1)^+_{k+1}$.

It becomes straightforward to write the 
two-excitation state as
\<
\state{\psi_1^\alpha\psi_2^\beta}\lvl{II}
\eq
\half\sum_{k=0}^K 
\Psi_{k}(y_1)\Psi_{k}(y_2)
\Bigl[
(\gen{Q}^{\alpha}{}_1)^-_k(\gen{Q}^{\beta}{}_1)^-_k
+2(\gen{Q}^{\alpha}{}_1)^-_k(\gen{Q}^{\beta}{}_1)^+_{k+1}
\nlnum\qquad\qquad\qquad\qquad\qquad\qquad\qquad\qquad
+(\gen{Q}^{\alpha}{}_1)^+_{k+1}(\gen{Q}^{\beta}{}_1)^+_{k+1}
 \bigr]\state{0}\lvl{II}
\nl
+\sum_{k<l=0}^K 
\Psi_{k}(y_1)\Psi_{l}(y_2)
\,
\bigbrk{(\gen{Q}^\alpha{}_1)^-_{k}+(\gen{Q}^\alpha{}_1)^+_{k+1}}
\bigbrk{(\gen{Q}^\beta{}_1)^-_{l}+(\gen{Q}^\beta{}_1)^+_{l+1}}
 \state{0}\lvl{II}.\nn
\>
Here, we have to make sure that the two excitations $y_1,y_2$ do
not cross when they are joined with the vacuum. This leads to the
slightly asymmetric form of the first term which should be understood 
as a chain-ordered version of the second term.
Now the two asymptotic regions are joined by 
\[
\state{\Psi^{\alpha\beta}_2}=
\state{\psi_1^\alpha\psi_2^\beta}\lvl{II}
+
\smat\lvl{II}_{12}
\state{\psi_1^\alpha\psi_2^\beta}\lvl{II}.
\]
There is no term for two coincident excitations anymore,
the correct factor in \eqref{eq:Diag.CoincidentFactor}
has been distributed among the two asymptotic regions.
The level-II S-matrix is 
\[
\smat\lvl{II}_{12}
\state{\psi_1^\alpha\psi_2^\beta}\lvl{II}=
M_{12}\state{\psi_2^\alpha\psi_1^\beta}\lvl{II}
+N_{12}\state{\psi_2^\beta\psi_1^\alpha}\lvl{II}.
\]
It should be clear how to generalise this framework to more than two excitations.

\section{A Singlet State}
\label{sec:Singlet}

In this appendix we construct and investigate a composite excitation 
which transforms as a singlet of the symmetry algebra.
We have no direct use for it, but its existence appears exciting.

\subsection{The State}

Considerations of the manifest symmetries $\alg{su}(3)$ and $\alg{su}(2)$ 
suggest that the singlet must be composed from the two 
building blocks 
$\varepsilon_{ab}\state{\phi^a_1\phi^b_2\fldZ^+\fldZ^+\fldZ^+}\lvl{I}$ and
$\varepsilon_{\alpha\beta}\state{\psi^\alpha_1\psi^\beta_2\fldZ^+\fldZ^+}\lvl{I}$.
To obtain the relative coefficient we demand invariance under
the fermionic generators in \eqref{eq:Rep.ActionQS}
and find
\[
\state{\rep{1}_{12}}\lvl{I}=
\frac{\alpha}{\gamma_1\gamma_2}\lrbrk{\frac{x_1^+}{x_1^-}-1}\varepsilon_{ab}
\state{\phi^a_1\phi^b_2\fldZ^+\fldZ^+\fldZ^+}\lvl{I}+
\varepsilon_{\alpha\beta}\state{\psi^\alpha_1\psi^\beta_2\fldZ^+\fldZ^+}\lvl{I}
\]
with $x_2^\pm=g^2/2x_1^\pm$. 
Also the central charges $\gen{C},\gen{P},\gen{K}$ annihilate this state.
It is clear that one of the excitations is not physical, 
it has a negative central charge which balances the 
positive central charge of the other excitation.
This composite of the two excitations might be interpreted as a
particle-antiparticle pair. One could also say that one of the
components is a creation operator while the 
other is an annihilation operator.

For $\superN=4$ SYM, the invariant combination
essentially consists of two covariant derivatives.
Their total anomalous dimension is $-2$ which cancels
precisely their contribution to the classical dimension.

\subsection{Scattering}

We can scatter the compound with any other excitation $\fldX$.
Remarkably, the compound stays intact and 
the scattering phase is independent of the type of excitation.
We find
\[\label{eq:Singlet.Scatter}
\smat\lvl{I}_{13}\smat\lvl{I}_{12}\state{\fldX_3\rep{1}_{12}}\lvl{I}=
\frac{x^+_3x^+_3}{x^-_3x^-_3}\,
\frac{x^+_3-x^+_1}{x^+_3-x^-_1}\,
\frac{1-g^2/2x^-_3x^+_1}{1-g^2/2x^-_3x^-_1}\,
S^0(x_3,x_1)\,S^0(x_3,g^2/2x_1)\,
\state{\rep{1}_{12}\fldX_3}\lvl{I}.
\]
Note that this is not symmetric under the map of
$x^\pm_1 \to g^2/2x^\pm_1$. 
This is okay as the compound $\rep{1}_{12}$ is 
not symmetric under the interchange of $x_1$ and $x_2$. 
In fact, trying to interchange the components of $\rep{1}_{12}$
is not well-defined due to divergencies in the S-matrix.


We can also represent the state by means of diagonalised excitations.
Then it is composed from the excitations 
$(K^0,K\lvl{I},K\lvl{II},K\lvl{III})=(2,2,2,1)$.
We can obtain trivial scattering for all
but the main excitations 
by setting $w=u$, $v_1=u+\ihalf$, $v_2=u-\ihalf$,
$x^\pm_2=g^2/2x^\pm_1$.
When we then set $y_1=x^+_1$ and $y_2=g^2/2x^-_1$ 
we obtain the same phase as in \eqref{eq:Singlet.Scatter}.

For the complete $\superN=4$ SYM model we can construct a similar
invariant state from the excitations
$(K^0,\dot K\lvl{III},\dot K\lvl{II},K\lvl{I},K\lvl{II},K\lvl{III})=(0,1,2,2,2,1)$.
Here, in addition we must set
$\dot w=u$, $\dot v_1=u+\ihalf$, $\dot v_2=u-\ihalf$.
For $\dot y_{1,2}$ we have to choose between $x_1^\pm$ and $g^2/2x_1^\pm$. 
Setting as above $\dot y_1=g^2/2x^+_1$ and $\dot y_2=x^-_1$
the overall phase is
\[
\smat\lvl{I}_{13}\smat\lvl{I}_{12}\state{\fldX_3\rep{1}_{12}}\lvl{I}=
\lrbrk{\frac{x^+_3-x^+_1}{x^+_3-x^-_1}\,
\frac{1-g^2/2x^-_3x^+_1}{1-g^2/2x^-_3x^-_1}}^2\,
S^0(x_3,x_1)\,S^0(x_3,g^2/2x_1)\,
\state{\rep{1}_{12}\fldX_3}\lvl{I}.
\]
This particular choice is most likely the correct one because the first term in the scattering
factor matches the function 
$f^2_{13}$ obtained in the context of crossing symmetry \cite{Janik:2006dc}.


\bibliography{n4smatrix}

\begin{thebibliography}{10}
\ifx\href\asklfhas\newcommand{\href}[2]{#2}\fi
\ifx\arxivref\asklfhas\newcommand{\arxivref}[1]{\href{http://arxiv.org/abs/#1}%
{#1}}\fi
\ifx\doiref\asklfhas\newcommand{\doiref}[2]{\href{http://dx.doi.org/#1}{#2}}\fi
\raggedright
\small
\parskip 0pt

\bibitem{Lipatov:1997vu}
L.~N.~Lipatov,
\textit{``Evolution equations in QCD''},
in: \textit{``Perspectives in hadronic physics''},
ed.: S.~Boffi, C.~Ciofi Degli~Atti and M.~Giannini,
World Scientific (1998),
Singapore.
%
\bibitem{Minahan:2002ve}
J.~A.~Minahan and K.~Zarembo,
\textit{``The Bethe-ansatz for {$\mathcal{N}=\mathord{}$4} super Yang-Mills''},
\textsf{JHEP~0303,~013~(2003)},
\texttt{\arxivref{hep-th/0212208}}.
%
\bibitem{Beisert:2003yb}
N.~Beisert and M.~Staudacher,
\textit{``The {$\mathcal{N}=\mathord{}$4} SYM Integrable Super Spin Chain''},
\textsf{\doiref{10.1016/j.nuclphysb.2003.08.015}{Nucl.~Phys.~B670,~439~(2003)}%
},
\texttt{\arxivref{hep-th/0307042}}.
%
\bibitem{Beisert:2003tq}
N.~Beisert, C.~Kristjansen and M.~Staudacher,
\textit{``The Dilatation Operator of {$\mathcal{N}=\mathord{}$4} Conformal
  Super Yang-Mills Theory''},
\textsf{\doiref{10.1016/S0550-3213(03)00406-1}{Nucl.~Phys.~B664,~131~(2003)}},
\texttt{\arxivref{hep-th/0303060}}.
%
\bibitem{Beisert:2003ys}
N.~Beisert,
\textit{``The SU(2$/$3) Dynamic Spin Chain''},
\textsf{\doiref{10.1016/j.nuclphysb.2003.12.032}{Nucl.~Phys.~B682,~487~(2004)}%
},
\texttt{\arxivref{hep-th/0310252}}.
%
\bibitem{Beisert:2004ry}
N.~Beisert,
\textit{``The Dilatation Operator of {$\mathcal{N}=\mathord{}$4} Super
  Yang-Mills Theory and Integrability''},
\textsf{\doiref{10.1016/j.physrep.2004.09.007}{Phys.~Rept.~405,~1~(2004)}},
\texttt{\arxivref{hep-th/0407277}}.
%
\bibitem{Beisert:2004yq}
N.~Beisert,
\textit{``Higher-Loop Integrability in {$\mathcal{N}=\mathord{}$4} Gauge
  Theory''},
\textsf{\doiref{10.1016/j.crhy.2004.09.011}{Comptes~Rendus~Physique~5,~1039~(2%
004)}},
\texttt{\arxivref{hep-th/0409147}}.
%
\bibitem{Zarembo:2004hp}
K.~Zarembo,
\textit{``Semiclassical Bethe ansatz and AdS/CFT''},
\textsf{Comptes~Rendus~Physique~5,~1081~(2004)},
\texttt{\arxivref{hep-th/0411191}}.
%
\bibitem{Plefka:2005bk}
J.~Plefka,
\textit{``Spinning strings and integrable spin chains in the AdS/CFT
  correspondence''},
\textsf{Living.~Rev.~Relativity~8,~9~(2005)},
\texttt{\arxivref{hep-th/0507136}}.
%
\bibitem{Eden:2004ua}
B.~Eden, C.~Jarczak and E.~Sokatchev,
\textit{``A three-loop test of the dilatation operator in
  {$\mathcal{N}=\mathord{}$4} SYM''},
\textsf{Nucl.~Phys.~B712,~157~(2005)},
\texttt{\arxivref{hep-th/0409009}}.
%
\bibitem{Inozemtsev:1989yq}
V.~I.~Inozemtsev,
\textit{``On the connection between the one-dimensional $s=1/2$ Heisenberg
  chain and Haldane Shastry model''},
\textsf{J.~Stat.~Phys.~59,~1143~(1990)}.
%
\bibitem{Inozemtsev:2002vb}
V.~I.~Inozemtsev,
\textit{``Integrable Heisenberg-van Vleck chains with variable range
  exchange''},
\textsf{Phys.~Part.~Nucl.~34,~166~(2003)},
\texttt{\arxivref{hep-th/0201001}}.
%
\bibitem{Serban:2004jf}
D.~Serban and M.~Staudacher,
\textit{``Planar {$\mathcal{N}=\mathord{}$4} gauge theory and the Inozemtsev
  long range spin chain''},
\textsf{JHEP~0406,~001~(2004)},
\texttt{\arxivref{hep-th/0401057}}.
%
\bibitem{Staudacher:2004tk}
M.~Staudacher,
\textit{``The factorized S-matrix of CFT/AdS''},
\textsf{JHEP~0505,~054~(2005)},
\texttt{\arxivref{hep-th/0412188}}.
%
\bibitem{Beisert:2004hm}
N.~Beisert, V.~Dippel and M.~Staudacher,
\textit{``A Novel Long Range Spin Chain and Planar {$\mathcal{N}=\mathord{}$4}
  Super Yang-Mills''},
\textsf{\doiref{10.1088/1126-6708/2004/07/075}{JHEP~0407,~075~(2004)}},
\texttt{\arxivref{hep-th/0405001}}.
%
\bibitem{Beisert:2005fw}
N.~Beisert and M.~Staudacher,
\textit{``Long-Range PSU(2,2$/$4) Bethe Ansaetze for Gauge Theory and
  Strings''},
\textsf{\doiref{10.1016/j.nuclphysb.2005.06.038}{Nucl.~Phys.~B727,~1~(2005)}},
\texttt{\arxivref{hep-th/0504190}}.
%
\bibitem{Maldacena:1998re}
J.~M.~Maldacena,
\textit{``The large N limit of superconformal field theories and
  supergravity''},
\textsf{Adv.~Theor.~Math.~Phys.~2,~231~(1998)},
\texttt{\arxivref{hep-th/9711200}}.
%
\bibitem{Gubser:1998bc}
S.~S.~Gubser, I.~R.~Klebanov and A.~M.~Polyakov,
\textit{``Gauge theory correlators from non-critical string theory''},
\textsf{Phys.~Lett.~B428,~105~(1998)},
\texttt{\arxivref{hep-th/9802109}}.
%
\bibitem{Witten:1998qj}
E.~Witten,
\textit{``Anti-de Sitter space and holography''},
\textsf{Adv.~Theor.~Math.~Phys.~2,~253~(1998)},
\texttt{\arxivref{hep-th/9802150}}.
%
\bibitem{Berenstein:2002jq}
D.~Berenstein, J.~M.~Maldacena and H.~Nastase,
\textit{``Strings in flat space and pp waves from {$\mathcal{N}=\mathord{}$4}
  {Super} {Yang Mills}''},
\textsf{JHEP~0204,~013~(2002)},
\texttt{\arxivref{hep-th/0202021}}.
%
\bibitem{Gubser:2002tv}
S.~S.~Gubser, I.~R.~Klebanov and A.~M.~Polyakov,
\textit{``A semi-classical limit of the gauge/string correspondence''},
\textsf{Nucl.~Phys.~B636,~99~(2002)},
\texttt{\arxivref{hep-th/0204051}}.
%
\bibitem{Frolov:2002av}
S.~Frolov and A.~A.~Tseytlin,
\textit{``Semiclassical quantization of rotating superstring in {$AdS_5 \times
  S^5$}''},
\textsf{JHEP~0206,~007~(2002)},
\texttt{\arxivref{hep-th/0204226}}.
%
\bibitem{Mandal:2002fs}
G.~Mandal, N.~V.~Suryanarayana and S.~R.~Wadia,
\textit{``Aspects of semiclassical strings in $AdS_5$''},
\textsf{Phys.~Lett.~B543,~81~(2002)},
\texttt{\arxivref{hep-th/0206103}}.
%
\bibitem{Bena:2003wd}
I.~Bena, J.~Polchinski and R.~Roiban,
\textit{``Hidden symmetries of the {$AdS_5\times S^5$} superstring''},
\textsf{Phys.~Rev.~D69,~046002~(2004)},
\texttt{\arxivref{hep-th/0305116}}.
%
\bibitem{Arutyunov:2004vx}
G.~Arutyunov, S.~Frolov and M.~Staudacher,
\textit{``Bethe ansatz for quantum strings''},
\textsf{JHEP~0410,~016~(2004)},
\texttt{\arxivref{hep-th/0406256}}.
%
\bibitem{Berkovits:2004xu}
N.~Berkovits,
\textit{``Quantum consistency of the superstring in $AdS_5\times S^5$
  background''},
\textsf{JHEP~0503,~041~(2005)},
\texttt{\arxivref{hep-th/0411170}}.
%
\bibitem{Frolov:2003qc}
S.~Frolov and A.~A.~Tseytlin,
\textit{``Multi-spin string solutions in {$AdS_5\times S^5$}''},
\textsf{Nucl.~Phys.~B668,~77~(2003)},
\texttt{\arxivref{hep-th/0304255}}.
%
\bibitem{Beisert:2003xu}
N.~Beisert, J.~A.~Minahan, M.~Staudacher and K.~Zarembo,
\textit{``Stringing Spins and Spinning Strings''},
\textsf{\doiref{10.1088/1126-6708/2003/09/010}{JHEP~0309,~010~(2003)}},
\texttt{\arxivref{hep-th/0306139}}.
%
\bibitem{Frolov:2003xy}
S.~Frolov and A.~A.~Tseytlin,
\textit{``Rotating string solutions: AdS/CFT duality in non-supersymmetric
  sectors''},
\textsf{Phys.~Lett.~B570,~96~(2003)},
\texttt{\arxivref{hep-th/0306143}}.
%
\bibitem{Beisert:2003ea}
N.~Beisert, S.~Frolov, M.~Staudacher and A.~A.~Tseytlin,
\textit{``Precision Spectroscopy of AdS/CFT''},
\textsf{\doiref{10.1088/1126-6708/2003/10/037}{JHEP~0310,~037~(2003)}},
\texttt{\arxivref{hep-th/0308117}}.
%
\bibitem{Callan:2003xr}
C.~G.~Callan,~Jr., H.~K.~Lee, T.~McLoughlin, J.~H.~Schwarz, I.~Swanson and
  X.~Wu,
\textit{``Quantizing string theory in $AdS_5\times S^5$: Beyond the pp-wave''},
\textsf{Nucl.~Phys.~B673,~3~(2003)},
\texttt{\arxivref{hep-th/0307032}}.
%
\bibitem{Callan:2004uv}
C.~G.~Callan,~Jr., T.~McLoughlin and I.~Swanson,
\textit{``Holography beyond the Penrose limit''},
\textsf{Nucl.~Phys.~B694,~115~(2004)},
\texttt{\arxivref{hep-th/0404007}}.
%
\bibitem{Tseytlin:2003ii}
A.~A.~Tseytlin,
\textit{``Spinning strings and AdS/CFT duality''},
\texttt{\arxivref{hep-th/0311139}},
in: \textit{``From fields to stings: Circumnavigating theoretical physics''},
ed.: M.~Shifman, A.~Vainshtein and J.~Wheater,
World Scientific (2005),
Singapore.
%
\bibitem{Swanson:2005wz}
I.~J.~Swanson,
\textit{``Superstring holography and integrability in $AdS_5\times S^5$''},
\texttt{\arxivref{hep-th/0505028}}.
%
\bibitem{Bethe:1931hc}
H.~Bethe,
\textit{``Zur Theorie der Metalle I. Eigenwerte und Eigenfunktionen der
  linearen Atomkette''},
\textsf{Z.~Phys.~71,~205~(1931)}.
%
\bibitem{Sutherland:1978aa}
B.~Sutherland,
\textit{``A brief history of the quantum soliton with new results on the
  quantization of the Toda lattice''},
\textsf{Rocky~Mountain~J.~Math.~8,~413~(1978)}.
%
\bibitem{Beisert:2005wm}
N.~Beisert,
\textit{``An SU(1$/$1)-Invariant S-Matrix with Dynamic Representations''},
\texttt{\arxivref{hep-th/0511013}},
in: \textit{``Quantum theory and symmetries''},
ed.: V.~K.~Dobrev,
Heron Press (2006),
Sofia, Bulgaria.
%
\bibitem{Yang:1967bm}
C.-N.~Yang,
\textit{``Some exact results for the many body problems in one dimension with
  repulsive delta function interaction''},
\textsf{Phys.~Rev.~Lett.~19,~1312~(1967)}.
%
\bibitem{Kotikov:2004er}
A.~V.~Kotikov, L.~N.~Lipatov, A.~I.~Onishchenko and V.~N.~Velizhanin,
\textit{``Three-loop universal anomalous dimension of the Wilson operators in
  {$\mathcal{N}=\mathord{}$4} SUSY Yang-Mills model''},
\textsf{Phys.~Lett.~B595,~521~(2004)},
\texttt{\arxivref{hep-th/0404092}}.
%
\bibitem{Moch:2004pa}
S.~Moch, J.~A.~M.~Vermaseren and A.~Vogt,
\textit{``The three-loop splitting functions in QCD: The non-singlet case''},
\textsf{Nucl.~Phys.~B688,~101~(2004)},
\texttt{\arxivref{hep-ph/0403192}}.
%
\bibitem{Vogt:2004mw}
A.~Vogt, S.~Moch and J.~A.~M.~Vermaseren,
\textit{``The three-loop splitting functions in QCD: The singlet case''},
\textsf{Nucl.~Phys.~B691,~129~(2004)},
\texttt{\arxivref{hep-ph/0404111}}.
%
\bibitem{Zamolodchikov:1978xm}
A.~B.~Zamolodchikov and A.~B.~Zamolodchikov,
\textit{``Factorized S-matrices in two dimensions as the exact solutions of
  certain relativistic quantum field models''},
\textsf{Annals~Phys.~120,~253~(1979)}.
%
\bibitem{Kazakov:2004qf}
V.~A.~Kazakov, A.~Marshakov, J.~A.~Minahan and K.~Zarembo,
\textit{``Classical/quantum integrability in AdS/CFT''},
\textsf{JHEP~0405,~024~(2004)},
\texttt{\arxivref{hep-th/0402207}}.
%
\bibitem{Kazakov:2004nh}
V.~A.~Kazakov and K.~Zarembo,
\textit{``Classical/quantum integrability in non-compact sector of AdS/CFT''},
\textsf{JHEP~0410,~060~(2004)},
\texttt{\arxivref{hep-th/0410105}}.
%
\bibitem{Beisert:2004ag}
N.~Beisert, V.~A.~Kazakov and K.~Sakai,
\textit{``Algebraic Curve for the SO(6) Sector of AdS/CFT''},
\textsf{\doiref{10.1007/s00220-005-1528-x}{Commun.~Math.~Phys.~263,~611~(2006)%
}},
\texttt{\arxivref{hep-th/0410253}}.
%
\bibitem{Schafer-Nameki:2004ik}
S.~Sch{\"a}fer-Nameki,
\textit{``The algebraic curve of 1-loop planar {$\mathcal{N}=\mathord{}$4}
  SYM''},
\textsf{Nucl.~Phys.~B714,~3~(2005)},
\texttt{\arxivref{hep-th/0412254}}.
%
\bibitem{Beisert:2005bm}
N.~Beisert, V.~Kazakov, K.~Sakai and K.~Zarembo,
\textit{``The Algebraic Curve of Classical Superstrings on $AdS_5\times
  S^5$''},
\textsf{\doiref{10.1007/s00220-006-1529-4}{Commun.~Math.~Phys.~263,~659~(2006)%
}},
\texttt{\arxivref{hep-th/0502226}}.
%
\bibitem{Das:2004hy}
A.~Das, J.~Maharana, A.~Melikyan and M.~Sato,
\textit{``The algebra of transition matrices for the $AdS_5\times S^5$
  superstring''},
\textsf{JHEP~0412,~055~(2004)},
\texttt{\arxivref{hep-th/0411200}}.
%
\bibitem{Arutyunov:2004yx}
G.~Arutyunov and S.~Frolov,
\textit{``Integrable Hamiltonian for classical strings on $AdS_5\times S^5$''},
\textsf{JHEP~0502,~059~(2005)},
\texttt{\arxivref{hep-th/0411089}}.
%
\bibitem{Alday:2005gi}
L.~F.~Alday, G.~Arutyunov and A.~A.~Tseytlin,
\textit{``On Integrability of Classical SuperStrings in $AdS_5\times S^5$''},
\textsf{JHEP~0507,~002~(2005)},
\texttt{\arxivref{hep-th/0502240}}.
%
\bibitem{Das:2005hp}
A.~Das, A.~Melikyan and M.~Sato,
\textit{``The algebra of flat currents for the string on $AdS_5\times S^5$ in
  the light-cone gauge''},
\textsf{JHEP~0511,~015~(2005)},
\texttt{\arxivref{hep-th/0508183}}.
%
\bibitem{Alday:2005jm}
L.~F.~Alday, G.~Arutyunov and S.~Frolov,
\textit{``New integrable system of 2dim fermions from strings on $AdS_5\times
  S^5$''},
\textsf{JHEP~0601,~078~(2006)},
\texttt{\arxivref{hep-th/0508140}}.
%
\bibitem{Arutyunov:2005hd}
G.~Arutyunov and S.~Frolov,
\textit{``Uniform light-cone gauge for strings in $AdS_5\times S^5$: Solving
  su(1$/$1) sector''},
\textsf{JHEP~0601,~055~(2006)},
\texttt{\arxivref{hep-th/0510208}}.
%
\bibitem{Frolov:2003tu}
S.~Frolov and A.~A.~Tseytlin,
\textit{``Quantizing three-spin string solution in {$AdS_5 \times S^5$}''},
\textsf{JHEP~0307,~016~(2003)},
\texttt{\arxivref{hep-th/0306130}}.
%
\bibitem{Frolov:2004bh}
S.~A.~Frolov, I.~Y.~Park and A.~A.~Tseytlin,
\textit{``On one-loop correction to energy of spinning strings in $S^5$''},
\textsf{Phys.~Rev.~D71,~026006~(2005)},
\texttt{\arxivref{hep-th/0408187}}.
%
\bibitem{Park:2005ji}
I.~Y.~Park, A.~Tirziu and A.~A.~Tseytlin,
\textit{``Spinning strings in $AdS_5\times S^5$: One-loop correction to energy
  in SL(2) sector''},
\textsf{JHEP~0503,~013~(2005)},
\texttt{\arxivref{hep-th/0501203}}.
%
\bibitem{Beisert:2005mq}
N.~Beisert, A.~A.~Tseytlin and K.~Zarembo,
\textit{``Matching Quantum Strings to Quantum Spins: One-Loop vs.~Finite-Size
  Corrections''},
\textsf{\doiref{10.1016/j.nuclphysb.2005.03.030}{Nucl.~Phys.~B715,~190~(2005)}%
},
\texttt{\arxivref{hep-th/0502173}}.
%
\bibitem{Hernandez:2005nf}
R.~Hern\'andez, E.~L\'opez, A.~Peri\'a\~nez and G.~Sierra,
\textit{``Finite size effects in ferromagnetic spin chains and quantum
  corrections to classical strings''},
\textsf{JHEP~0506,~011~(2005)},
\texttt{\arxivref{hep-th/0502188}}.
%
\bibitem{Beisert:2005bv}
N.~Beisert and L.~Freyhult,
\textit{``Fluctuations and Energy Shifts in the Bethe Ansatz''},
\textsf{\doiref{10.1016/j.physletb.2005.07.015}{Phys.~Lett.~B622,~343~(2005)}},
\texttt{\arxivref{hep-th/0506243}}.
%
\bibitem{Schafer-Nameki:2005tn}
S.~Sch{\"a}fer-Nameki, M.~Zamaklar and K.~Zarembo,
\textit{``Quantum corrections to spinning strings in $AdS_5\times S^5$ and
  Bethe ansatz: A comparative study''},
\textsf{JHEP~0509,~051~(2005)},
\texttt{\arxivref{hep-th/0507189}}.
%
\bibitem{Beisert:2005cw}
N.~Beisert and A.~A.~Tseytlin,
\textit{``On Quantum Corrections to Spinning Strings and Bethe Equations''},
\textsf{\doiref{10.1016/j.physletb.2005.09.054}{Phys.~Lett.~B629,~102~(2005)}},
\texttt{\arxivref{hep-th/0509084}}.
%
\bibitem{Schafer-Nameki:2005is}
S.~Sch{\"a}fer-Nameki and M.~Zamaklar,
\textit{``Stringy sums and corrections to the quantum string Bethe ansatz''},
\textsf{JHEP~0510,~044~(2005)},
\texttt{\arxivref{hep-th/0509096}}.
%
\bibitem{Gromov:2005gp}
N.~Gromov and V.~Kazakov,
\textit{``Double scaling and finite size corrections in SL(2) spin chain''},
\textsf{Nucl.~Phys.~B736,~199~(2006)},
\texttt{\arxivref{hep-th/0510194}}.
%
\bibitem{Mann:2004jr}
N.~Mann and J.~Polchinski,
\textit{``Finite density states in integrable conformal field theories''},
\texttt{\arxivref{hep-th/0408162}},
in: \textit{``From fields to stings: Circumnavigating theoretical physics''},
ed.: M.~Shifman, A.~Vainshtein and J.~Wheater,
World Scientific (2005),
Singapore.
%
\bibitem{Mann:2005ab}
N.~Mann and J.~Polchinski,
\textit{``Bethe ansatz for a quantum supercoset sigma model''},
\textsf{Phys.~Rev.~D72,~086002~(2005)},
\texttt{\arxivref{hep-th/0508232}}.
%
\bibitem{Lin:2005nh}
H.~Lin and J.~Maldacena,
\textit{``Fivebranes from gauge theory''},
\texttt{\arxivref{hep-th/0509235}}.
%
\bibitem{Kim:2003rz}
N.~Kim, T.~Klose and J.~Plefka,
\textit{``Plane-wave Matrix Theory from {$\mathcal{N}=\mathord{}$4} Super
  Yang-Mills on {$R\times S^3$}''},
\textsf{Nucl.~Phys.~B671,~359~(2003)},
\texttt{\arxivref{hep-th/0306054}}.
%
\bibitem{Fischbacher:2004iu}
T.~Fischbacher, T.~Klose and J.~Plefka,
\textit{``Planar plane-wave matrix theory at the four loop order: Integrability
  without BMN scaling''},
\textsf{JHEP~0502,~039~(2005)},
\texttt{\arxivref{hep-th/0412331}}.
%
\bibitem{Klose:2005aa}
T.~Klose,
\textit{``Plane Wave Matrix Theory: Gauge Theoretical Origin and Planar
  Integrability''},
PhD thesis.
%
\bibitem{Beisert:2005wv}
N.~Beisert and T.~Klose,
\textit{``Long-Range $GL(n)$ Integrable Spin Chains and Plane-Wave Matrix
  Theory''},
\textsf{\doiref{10.1088/1742-5468/2006/07/P07006}{J.~Stat.~Mech.~06,~P07006~(2%
006)}},
\texttt{\arxivref{hep-th/0510124}}.
%
\bibitem{Zwiebel:2005er}
B.~I.~Zwiebel,
\textit{``{$\mathcal{N}=\mathord{}$4} SYM to two loops: Compact expressions for
  the non-compact symmetry algebra of the su(1,1$/$2) sector''},
\textsf{JHEP~0602,~055~(2006)},
\texttt{\arxivref{hep-th/0511109}}.
%
\bibitem{Ambjorn:2005wa}
J.~Ambj{\o}rn, R.~A.~Janik and C.~Kristjansen,
\textit{``Wrapping interactions and a new source of corrections to the
  spin-chain / string duality''},
\textsf{Nucl.~Phys.~B736,~288~(2006)},
\texttt{\arxivref{hep-th/0510171}}.
%
\bibitem{Iohara:2001aa}
K.~Iohara and Y.~Koga,
\textit{``Central extensions of Lie superalgebras''},
\textsf{Comment.~Math.~Helv.~76,~110~(2001)}.
%
\bibitem{Essler:1992aa}
F.~H.~L.~E{\ss}ler, V.~E.~Korepin and K.~Schoutens,
\textit{``New Exactly Solvable Model of Strongly Correlated Electrons Motivated
  by High-$T_c$ Superconductivity''},
\textsf{Phys.~Rev.~Lett.~68,~2960~(1992)},
\texttt{\arxivref{cond-mat/9209002}}.
%
\bibitem{Santambrogio:2002sb}
A.~Santambrogio and D.~Zanon,
\textit{``Exact anomalous dimensions of {$\mathcal{N}=\mathord{}$4} Yang-Mills
  operators with large R charge''},
\textsf{Phys.~Lett.~B545,~425~(2002)},
\texttt{\arxivref{hep-th/0206079}}.
%
\bibitem{Minahan:2005jq}
J.~A.~Minahan,
\textit{``The SU(2) sector in AdS/CFT''},
\textsf{Fortsch.~Phys.~53,~828~(2005)},
\texttt{\arxivref{hep-th/0503143}}.
%
\bibitem{Agarwal:2005jj}
A.~Agarwal and G.~Ferretti,
\textit{``Higher charges in dynamical spin chains for SYM theory''},
\textsf{JHEP~0510,~051~(2005)},
\texttt{\arxivref{hep-th/0508138}}.
%
\bibitem{Pope:2003jp}
C.~N.~Pope and N.~P.~Warner,
\textit{``A dielectric flow solution with maximal supersymmetry''},
\textsf{JHEP~0404,~011~(2004)},
\texttt{\arxivref{hep-th/0304132}}.
%
\bibitem{Bena:2004jw}
I.~Bena and N.~P.~Warner,
\textit{``A harmonic family of dielectric flow solutions with maximal
  supersymmetry''},
\textsf{JHEP~0412,~021~(2004)},
\texttt{\arxivref{hep-th/0406145}}.
%
\bibitem{Bracken:1994hz}
A.~J.~Bracken, M.~D.~Gould, Y.-Z.~Zhang and G.~W.~Delius,
\textit{``Solutions of the quantum Yang-Baxter equation with extra nonadditive
  parameters''},
\textsf{J.~Phys.~A27,~6551~(1994)},
\texttt{\arxivref{hep-th/9405138}}.
%
\bibitem{Bracken:1995aa}
A.~J.~Bracken, M.~D.~Gould, J.~R.~Links and Y.-Z.~Zhang,
\textit{``New Supersymmetric and Exactly Solvable Model of Correlated
  Electrons''},
\textsf{Phys.~Rev.~Lett.~75,~2768~(1995)},
\texttt{\arxivref{cond-mat/9410026}}.
%
\bibitem{Pfannmuller:1996vp}
M.~P.~Pfannm{\"u}ller and H.~Frahm,
\textit{``Algebraic Bethe Ansatz for gl(2,1) Invariant 36-Vertex Models''},
\textsf{Nucl.~Phys.~B479,~575~(1996)},
\texttt{\arxivref{cond-mat/9604082}}.
%
\bibitem{Ramos:1996my}
P.~B.~Ramos and M.~J.~Martins,
\textit{``One parameter family of an integrable spl(2$/$1) vertex model:
  Algebraic Bethe ansatz approach and ground state structure''},
\textsf{Nucl.~Phys.~B474,~678~(1996)},
\texttt{\arxivref{hep-th/9604072}}.
%
\bibitem{Janik:2006dc}
R.~A.~Janik,
\textit{``The $AdS_5\times S^5$ superstring worldsheet S-matrix and crossing
  symmetry''},
\textsf{Phys.~Rev.~D73,~086006~(2006)},
\texttt{\arxivref{hep-th/0603038}}.
%
\end{thebibliography}
\bibliographystyle{nb}

\end{document}